\documentclass[12pt,preprint]{aastex}
\makeatletter
\renewcommand{\fps@table}{t}

\makeatother
\newcommand{\nodatr}{\multicolumn{1}{r}{$\cdots$}~~}
\newcommand{\kms}{\nobreak\mbox{$\;$km\,s$^{-1}$}}
\renewcommand{\mag}{\mbox{$\;$mag}}

\shorttitle{The Hubble Diagram of SNeIa within $30\,000\kms$}
\shortauthors{Reindl et al.}

\begin{document}
\title{REDDENING, ABSORPTION, AND DECLINE RATE CORRECTIONS FOR A 
       COMPLETE SAMPLE OF TYPE Ia SUPERNOVAE LEADING TO A FULLY 
       CORRECTED HUBBLE DIAGRAM TO \boldmath{$v < 30\,000\kms$}} 
\author{B. Reindl and G. A. Tammann}
\affil{Astronomisches Institut der Universit\"at Basel,\\
       Venusstrasse 7, CH-4102 Binningen, Switzerland}
\email{G-A.Tammann@unibas.ch}
\author{A. Sandage}
\affil{The Observatories of the Carnegie Institution of Washington,\\
       813 Santa Barbara Street, Pasadena, CA 91101}
\and
\author{A. Saha}
\affil{National Optical Astronomy Observatories,\\
       950 North Cherry Ave., Tucson, AZ 85726}

\begin{abstract}
Photometric ($BVI$) and redshift data corrected for streaming 
motions are compiled for 111 ``Branch normal'', four 1991T-like,
seven 1991bg-like, and two unusual  supernovae of type Ia. Color
excesses $E(B\!-\!V)_{\rm host}$ of normal SNe\,Ia, due to the
absorption of the host galaxy, are derived by three independent
methods, giving excellent agreement leading to the intrinsic colors at
maximum of $(B\!-\!V)^{00} = -0.024\pm0.010$, and $(V\!-\!I)^{00} =
-0.265\pm0.016$ if normalized to a common decline rate of $\Delta
m_{15} = 1.1$.    

     The strong correlation between redshift absolute 
magnitudes (based on an arbitrary Hubble constant of
$H_{0}=60\;\mbox{km}\,\mbox{s}^{-1}\,\mbox{Mpc}^{-1}$), corrected  
only for the extrinsic Galactic absorption, and the derived 
$E(B\!-\!V)_{\rm host}$ color excesses leads to the well determined,
yet abnormal absorption-to-reddening ratios of 
${\cal R}_{BVI}=3.65\pm0.16$, $2.65\pm0.15$, and $1.35\pm0.21$. 
Comparison with the canonical Galactic values of 
$4.1$, $3.1$, $1.8$ forces the conclusion that the law of interstellar 
absorption in the path length to the SN in the host galaxy is
different from the local Galactic law, a result
consistent with earlier conclusions by others. 
 
     Improved correlations of the fully corrected absolute magnitudes 
(on the same arbitrary Hubble constant zero point) with host 
galaxy morphological type, decline rate, and intrinsic color are 
derived. We recover the result that SN\,Ia in E/S0 galaxies are   
$\sim\!0.3$ magnitudes fainter than in spirals for possible reasons 
discussed in the text. The new decline-rate corrections to 
absolute magnitudes are smaller than those by some authors for reasons
explained in the text.

     The four spectroscopically peculiar 1991T-type SNe are 
significantly overluminous as compared to Branch-normal SNe\,Ia. The
overluminosity of the seven 1999aa-like SNe is less pronounced. The
seven 1991bg-types in the sample constitute a separate class of
SNe\,Ia, averaging in $B$ two magnitudes fainter than the normal Ia.      

     New Hubble diagrams in $B$, $V$, and $I$ are derived out to
$\sim\!30\,000\kms$ using the fully corrected magnitudes and
velocities, corrected for streaming motions.
Nine solutions for the intercept magnitudes in these diagrams show
extreme stability at the $0.04\mag$ level using various
subsamples of the data for both low and high extinctions in the
sample, proving the validity of the corrections for host 
galaxy absorption. -- The same precepts for fully correcting SN
magnitudes we shall use for the luminosity recalibration of SNe\,Ia
in the forthcoming final review of our HST
Cepheid-SN experiment for the Hubble constant.  
\end{abstract}
\keywords{distance scale --- galaxies: distances and redshifts ---
          supernovae: general}

\section{INTRODUCTION}
This is the third paper of a series whose purpose is to 
prepare for the projected final summary review of our 1991-2001 
Hubble Space Telescope experiment to determine the Hubble 
constant by calibrating the absolute magnitude of type Ia 
supernovae. The method used in this observing program has been to 
measure Cepheid distances to the host galaxies by 
(1) correcting the raw Cepheid photometric data for reddening to
determine the true distance moduli of the parent galaxies using an
adopted Cepheid period-luminosity relation and 
(2) correcting the SNe data for their own intrinsic and extrinsic
reddening and absorption values, plus all known second-parameter
corrections for light-curve inhomogeneities.  

     The first two papers of this preparatory series concerned 
revisions in the Cepheid P-L relation and its lack 
of universality from galaxy to galaxy, showing significant 
differences in the P-L slope and zero point between the Galaxy 
\citep{Tammann:etal:03} and the LMC \citep{Sandage:etal:04}.  

     Following the emergence of SNe\,Ia as powerful standard 
candles \citep{Kowal:68}, it became increasingly clear as the accuracy
of the data became better that the dispersion in the SNe\,Ia
absolute magnitude at maximum became smaller with each advance in 
the accuracy of the data 
\citep{Sandage:Tammann:82,Sandage:Tammann:93,Sandage:Tammann:97,{Saha:etal:97}},
and also with the application of corrections for light-curve
inhomogeneities that had been found and  
subsequently systematically improved
(\citealt{Pskovskii:67,Pskovskii:71,Pskovskii:84}; 
 \citealt{Barbon:etal:73};
 \citealt{Phillips:93};
 \citealt{Sandage:Tammann:95}; 
 \citealt{Hamuy:etal:95,Hamuy:etal:96a,Hamuy:etal:96d}; 
 \citealt{Tripp:98};
 \citealt{Phillips:etal:99};
 \citealt{Riess:etal:99};
 \citealt{Jha:etal:99};
 \citealt{Tripp:Branch:99};
 \citealt{Parodi:etal:00};
 \citealt{Tonry:etal:03}; 
 also \citealt{Sandage:etal:01} for a review).
             
     The purpose of this third paper of the series is to update 
the discussion by \citet{Parodi:etal:00} on reddening, 
absorption, and second parameter corrections (decline rate, 
color, galaxy type) to the SNe\,Ia light curves, and to determine 
a revised Hubble diagram based on these corrections using the 
totality of the modern SNe\,Ia photometry available to date. The 
sample here consists of 124 supernovae compared with the sample 
of 35 SNe\,Ia used by \citeauthor{Parodi:etal:00}.    

     The organization of the paper is this: 

     (1) The photometric data for 124 SNe\,Ia are compiled in 
\S~\ref{sec:Data} from the extensive literature, together with
the types and recession velocities of their host galaxies
(Table~\ref{tab:SN1}).  

     (2) The {\em intrinsic\/} colors $(B\!-\!V)^{00}_{\max}$,
$(B\!-\!V)^{00}_{35}$ (i.e.\ 35 days after $B$ maximum), and
$(V\!-\!I)^{00}_{\max}$ are derived in \S~\ref{sec:color} from 34
normal SNe\,Ia which have occurred in E/S0 galaxies or in the outer
regions of spiral galaxies and which are assumed -- after correction
for Galactic reddening -- to suffer no reddening in their host
galaxies. 

     (3) After correction for Galactic reddening, the reddening of the
normal SNe\,Ia due to the host galaxy is derived by three independent
methods in \S~\ref{sec:hostgal}, leading to the intrinsic (unreddened)
colors of each of the SNe. These corrected data are listed in
Table~\ref{tab:SN3}a.

     (4) New absorption-to-reddening ratios for that part of the 
reddening due to the host galaxy are derived in
\S~\ref{sec:absorption}, leading to the absorption corrections to
the observed SN apparent magnitudes (also corrected for the K-term
effect due to redshift). The absorption-free magnitudes are also
listed in Table~\ref{tab:SN3}a.  

     (5) New dependencies of the fully corrected SN magnitudes on 
galaxy type, decline rate $\Delta m_{15}$, and intrinsic color are 
derived in \S~\ref{sec:other}. Magnitudes $m_{BVI}^{\rm corr}$,
corrected for absorption and normalized to $\Delta m_{15}=1.1$ and
$(B\!-\!V)^{00}_{\max}=-0.024$ complete Table~\ref{tab:SN3}a. 

    (6)  A discussion is made in \S~\ref{sec:SNpec} of the
photometric properties of the small group (15\% of the total sample)
of peculiar SNe\,Ia, i.e.\ SNe-91T, SNe-99aa, SNe-91bg (named after
their eponymous prototypes), and two singular objects (SN-2000cx,
SN-1986G), with the conclusion that SNe-91T, including SNe-99aa, are
brighter than normal SNe\,Ia by $0.3$ and $0.2\mag$, respectively,
even if normalized to $\Delta m_{15}=1.1$, while SNe-91bg are
underluminous in $B$ by $2\mag$.             

    (7) Hubble diagrams using $m_{BVI}^{\rm corr}$ are derived for
various subsamples of normal SNe\,Ia in \S~\ref{sec:hubble}, together
with the rms magnitude deviations for each of the subsamples used. The
robustness of the solutions is emphasized.     

    (8) Twelve principal research points made in the paper 
are summarized in \S~\ref{sec:conclusions}.  

     The conventions used here are these: 
Colors $(X\!-\!Y)_{\max}$ are always $(X_{\max} - Y_{\max})$. 
$(X\!-\!Y)^{0}$ stands for colors corrected for Galactic reddening,
$(X\!-\!Y)^{00}$ for colors corrected for Galactic reddening 
{\em and\/} reddening in the host galaxy. 
Absolute magnitudes $M_{BVI}$ and all listed distance moduli are
calculated from the redshift using a flat universe with 
$\Omega_{\rm matter} = 0.3$, $\Omega_{\Lambda} = 0.7$ with an 
arbitrarily assumed Hubble constant of
$H_{0}=60\;\mbox{km}\,\mbox{s}^{-1}\,\mbox{Mpc}^{-1}$. None of 
the conclusions as to magnitude corrections and trends of the 
second-parameter corrections depend on $H_{0}$, hence there is no loss
of generality using any assumed value.

\section{THE DATA}
\label{sec:Data}
A list of 19 ``local'' ($v_{220} < 2000\kms$) and 94  ``nearby''
($v_{\rm CMB}\la 30\,000\kms$) supernovae of Type Ia (SNe\,Ia) after
1985 with available maximum magnitudes in $B$ and $V$, and, if
possible in $I$, is given in Table~\ref{tab:SN1}. 
The list aims to be complete as to published template-fitted maximum
magnitudes in $B$ and $V$, however the few nearby SNe\,Ia with not so
well observed lightcurves by
\citet{Germany:etal:04} are not considered. Eleven earlier 
SNe\,Ia (9 ``local'' and 2 ``nearby'') with useful photometry are
added.  
Table~\ref{tab:SN1} is subdivided into a) 111 spectroscopically
``normal'' SNe\,Ia in the sense of \citet{Branch:etal:93} including
objects which are not spectroscopically confirmed but share the
photometric properties of normal SNe\,Ia (which is assumed here in
first approximation also for the seven 1999aa-like objects); b) the
four SNe-91T which are spectroscopically peculiar at early phases
\citep{Saha:etal:01,Li:etal:01b}; in addition included is the
unique object SN\,2000cx whose spectrum resembles SN\,1991T at least
near maximum phase \citep{Li:etal:01c,Candia:etal:03}; and c) the
seven SNe-91bg objects which constitute a homogeneous sub-class
with very fast decline rates ($\Delta m_{15}\approx1.9$), and which
are quite red and underluminous at maximum. Also included here is
SN\,1986G; it has, however, a faster decline rate and much of its
apparent redness is probably due to absorption in the host galaxy.

Table~\ref{tab:SN1} is organized as follows:\\
Col.~1: The name of the SN\\
Col.~2: The name of the host galaxy\\
Col.~3: The coded type of the host galaxy: E$ = -3$, E/S0$ = -2$, S0$ = - 1$,
        S0/a$ = 0$, the later types as in
        \citet{deVaucouleurs:etal:76}. The galaxy types are taken from
        the same sources as the photometry, in some cases from
        \citet{vandenBergh:etal:02,vandenBergh:etal:03}. \\
Col.~4: The mean heliocentric radial velocity as given in the 
        NASA Extragalactic Database (NED). \\
Col.~5: The radial velocity corrected for streaming motions. The
        nearer galaxies are corrected for a self-consistent
        Virgocentric infall model with a local infall vector of
        $220\kms$ \citep{Kraan-Korteweg:86}. The more distant galaxies are 
        corrected for the CMB motion, where it is assumed that the
        co-moving volume extends to $v_{220} = 3000\kms$, justified by
        the kinematics of the ``local bubble'' that merges into the
        background field kinematics near $3000\kms$
        \citep{Federspiel:etal:94}. Even if merging takes place as far
        out as $6000\kms$ \citep{Dale:Giovanelli:00} it has no
        noticeable effect on the present results.
        If $v_{220} < 3000\kms$, the value of $v_{220}$ is listed
        (denoted with v). All larger velocities are in the CMB
        frame. 
        Virgo and Fornax cluster members are listed with the mean
        cluster velocity (denoted with V and F). 
        Also two galaxies in the W-Cloud (NGC\,4527, 4536) outside the
        Virgo cluster proper, are tentatively assigned 
        the cluster velocity.\\    
Col.~6-8: Observed $B_{\max}$, $V_{\max}$, and $I_{\max}$ magnitudes
        as compiled from 
        \citet{Hamuy:etal:96b,Hamuy:etal:96d}, \citet{Riess:etal:99},
        and \citet{Jha:02}. The latter has published  the $V$ and $I$
        magnitude at the epoch of the $B$ maximum; they were
        transformed  to $V_{\max}$ and $I_{\max}$ by interpolating
        Hamuy's et al. (1996c) template light curves (their 
        Table~10) with the appropriate decline rate $\Delta m_{15}$. 
        \citet{Riess:etal:99} have published only $B$ and $V$ maxima,
        their $I$ photometry was reduced by \citet{Parodi:etal:00} by
        the template method of \citet{Hamuy:etal:96c} -- which is also
        the procedure that provides the K-corrections for the effects
        of redshift on the photometry and which is the basis of all
        magnitudes in this paper. Where available the $B_{\max}$
        were taken from \citet{Altavilla:etal:04} who have uniformly
        reduced the available data including various other
        sources. The corresponding $V_{\max}$ are not published, but
        were kindly provided by Dr.~G.~Altavilla. Several additional
        sources are quoted in column~14.  \\
Col.~9: The decline rate $\Delta m_{15}$ taken from the same sources
as the apparent maximum magnitudes. The slight dependence of $\Delta
m_{15}$ on the reddening \citep{Phillips:etal:99} is neglected. \\
Col.~10: The Galactic reddening following \citet{Schlegel:etal:98}\\
Col.~11: The reddening $E(B\!-\!V)_{\max}$ in the host galaxy as
        judged from the $(B\!-\!V)^{0}$ color at maximum assuming
        intrinsic colors of $(B\!-\!V)^{00}_{\max}$ as given in
        equation~(\ref{eq:color:mean:BV}). \\ 
Col.~12: The reddening in the host galaxy at maximum as
        judged from the observed color $(B\!-\!V)^{00}_{35}$ 35
        days past $B$ maximum as explained in \S~\ref{sec:hostgal:BVmax}. \\ 
Col.~13: The reddening in the host galaxy as judged from the
        $(V\!-\!I)^{0}$ color at maximum assuming intrinsic colors
        $(V\!-\!I)^{00}_{\max}$ as given in
        equation~(\ref{eq:color:mean:VI}).   \\ 
Col.~14: The source(s) of the photometry. \\

\section{THE INTRINSIC COLOR OF SNe\,Ia}
\label{sec:color}
\begin{figure*}[t]
     \epsscale{0.9}
     \plotone{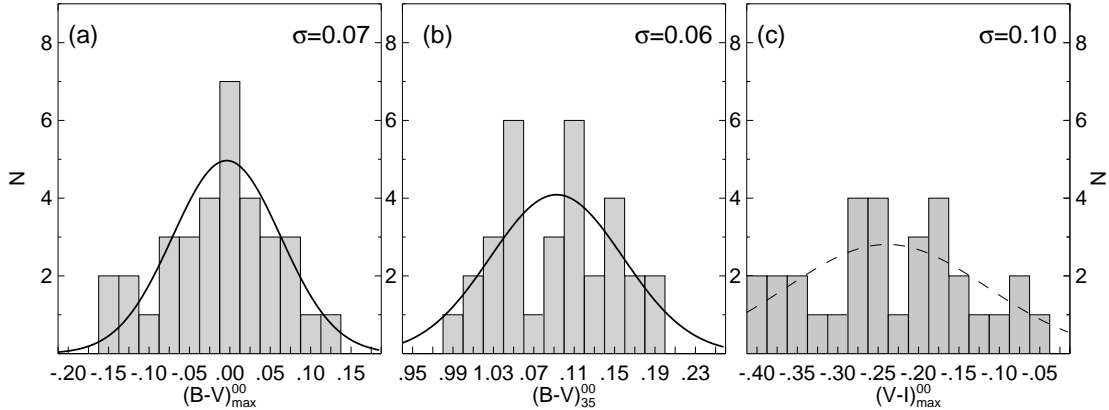}
     \caption{The true color distribution in $(B\!-\!V)_{\max}^{00}$,
     $(B\!-\!V)_{35}^{00}$, and $(V\!-\!I)_{\max}^{00}$ of SNe\,Ia
     with minimum reddening in their host galaxies from
     Table~\ref{tab:color:minred}.}   
\label{fig:color:tab2}
\end{figure*}

\begin{figure*}[t]
     \plotone{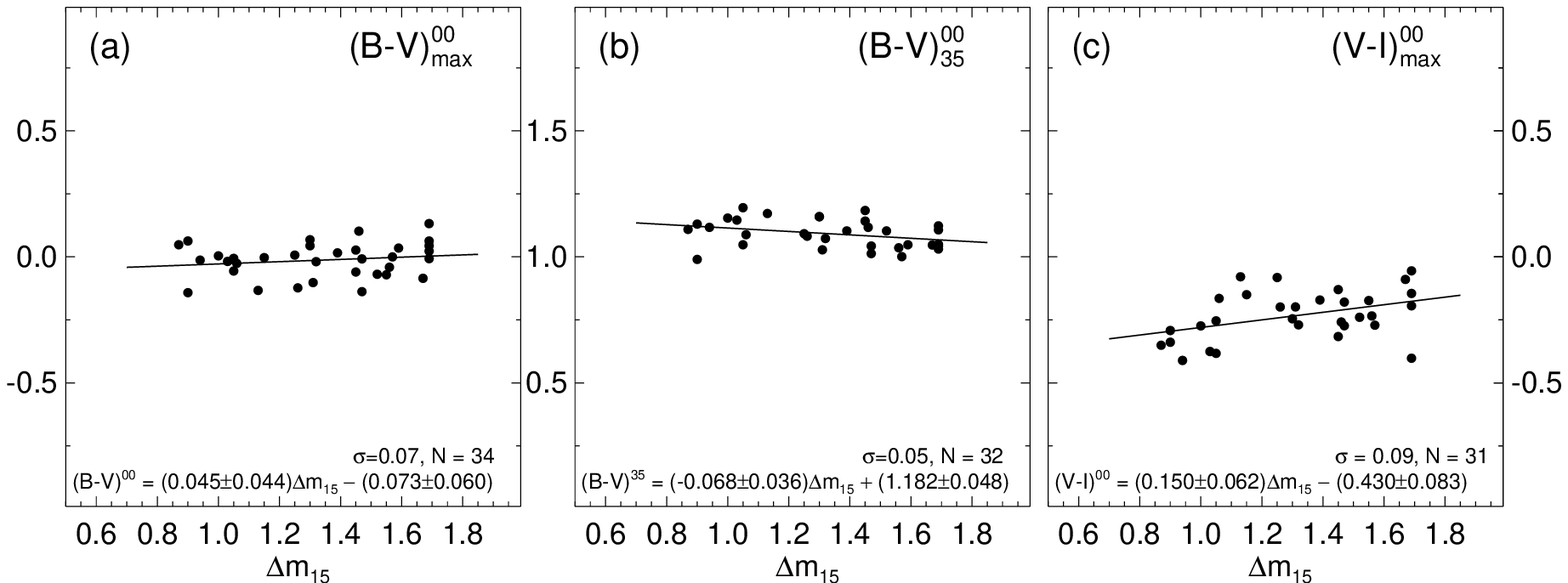}
     \caption{The dependence of the colors $(B\!-\!V)_{\max}^{00}$,
     $(B\!-\!V)_{35}^{00}$, and $(V\!-\!I)_{\max}^{00}$ of the SNe\,Ia
     in Table~\ref{tab:color:minred} on the decline rate $\Delta
     m_{15}$.}    
\label{fig:color:tab2dm15}
\end{figure*}

It is expected that SNe in early-type galaxies (S0 and earlier) as
well as far outlying SNe in spiral galaxies suffer negligible
reddening. In the case of SNe\,Ia in spiral galaxies we require that
their distance $r$ from the center of the host galaxy be larger than
$0.4r_{25}$, where $r_{25}$ is the isophote corresponding to 
$25\mag$ per arcsec$^{2}$ \citep[cf.][]{Parodi:etal:00}. A complete
list of 34 such SNe\,Ia is given in Table~\ref{tab:color:minred}
(omitting only the very red SN\,1974G). 
Their colors $(B\!-\!V)^{0}$, $(B\!-\!V)^{0}_{35}$, i.e.\ the color 35
days past $B$ maximum from \S~\ref{sec:hostgal:BVmax} below, 
and $(V\!-\!I)^{0}$, all corrected  for 
Galactic reddening (taken from column~10 of Table~\ref{tab:SN1}), are
listed in Table~\ref{tab:color:minred}. 
For the derivation of $(V\!-\!I)^{0}$ it was assumed that 
$E(V\!-\!I)_{\rm Gal} = 1.3E(B\!-\!V)_{\rm Gal}$. 
The distribution of the colors is shown in
Figure~\ref{fig:color:tab2}, their dependence on the decline rate
$\Delta m_{15}$ in Figure~\ref{fig:color:tab2dm15}.
\begin{deluxetable}{lrrrrrrr}
\def\baselinestretch{1.1}
\tablewidth{0pt}
\tabletypesize{\scriptsize}
\tablenum{2}
\tablecaption{SNe\,Ia with minimum reddening.\label{tab:color:minred}}
\tablehead{
\colhead{SN} & 
\colhead{T}  & 
\colhead{$(B\!-\!V)_{\max}$} & 
\colhead{$E(B\!-\!V)_{\rm Gal}$} &
\colhead{$(B\!-\!V)_{\max}^{0}$} & 
\colhead{$(B\!-\!V)_{35}^{0}$} & 
\colhead{$(V\!-\!I)_{\max}^{0}$} & 
\colhead{$\Delta m_{15}$}
\\ 
\colhead{(1)} & \colhead{(2)} & \colhead{(3)} & \colhead{(4)} & 
\colhead{(5)} & \colhead{(6)} & \colhead{(7)} & \colhead{(8)}
}
\startdata
\multicolumn{8}{c}{(a) SNe in early-type galaxies}\\
\noalign{\smallskip}
\tableline
1990af  & $-1$ & $ 0.07$ & $0.035$ & $ 0.035$ & $1.048$ & \nodatr  & $1.59$ \\
1992J   & $-2$ & $ 0.12$ & $0.057$ & $ 0.063$ & $1.107$ & $-0.194$ & $1.69$ \\
1992ae  & $-3$ & $ 0.08$ & $0.036$ & $ 0.044$ & $1.159$ & \nodatr  & $1.30$ \\
1992au  & $-3$ & $ 0.04$ & $0.017$ & $ 0.023$ & $1.050$ & $-0.402$ & $1.69$ \\
1992bk  & $-3$ & $-0.07$ & $0.015$ & $-0.085$ & $1.047$ & $-0.090$ & $1.67$ \\
1992bo  & $-2$ & $ 0.02$ & $0.027$ & $-0.007$ & $1.031$ & $-0.145$ & $1.69$ \\
1992bp  & $-2$ & $ 0.00$ & $0.069$ & $-0.069$ & $1.103$ & $-0.240$ & $1.52$ \\
1992br  & $-3$ & $ 0.07$ & $0.026$ & $ 0.044$ & $1.043$ & \nodatr  & $1.69$ \\
1993O   & $-2$ & $-0.07$ & $0.053$ & $-0.123$ & $1.082$ & $-0.199$ & $1.26$ \\
1993ac  & $-3$ & $ 0.17$ & $0.163$ & $ 0.007$ & $1.092$ & $-0.082$ & $1.25$ \\
1993ae  & $-3$ & $-0.10$ & $0.038$ & $-0.138$ & $1.013$ & $-0.179$ & $1.47$ \\
1993ag  & $-2$ & $ 0.18$ & $0.112$ & $ 0.068$ & $1.160$ & $-0.246$ & $1.30$ \\
1993ah  & $-1$ & $-0.04$ & $0.020$ & $-0.060$ & $1.142$ & $-0.316$ & $1.45$ \\
1994D   & $-1$ & $-0.08$ & $0.022$ & $-0.102$ & $1.028$ & $-0.199$ & $1.31$ \\
1994M   & $-3$ & $ 0.05$ & $0.023$ & $ 0.027$ & $1.184$ & $-0.130$ & $1.45$ \\
1994Q   & $-1$ & $ 0.08$ & $0.017$ & $ 0.063$ & $1.130$ & $-0.292$ & $0.90$ \\
1996X   & $-3$ & $ 0.05$ & $0.069$ & $-0.019$ & $1.073$ & $-0.270$ & $1.32$ \\
1997E   & $-1$ & $ 0.14$ & $0.124$ & $ 0.016$ & $1.103$ & $-0.171$ & $1.39$ \\
1998dx  & $-3$ & $-0.03$ & $0.041$ & $-0.071$ & \nodatr & $-0.173$ & $1.55$ \\
1999gh  & $-3$ & $ 0.19$ & $0.058$ & $ 0.132$ & $1.123$ & $-0.055$ & $1.69$ \\
2000B   & $-3$ & $ 0.17$ & $0.068$ & $ 0.102$ & $1.117$ & $-0.258$ & $1.46$ \\
2000dk  & $-3$ & $ 0.07$ & $0.070$ & $ 0.000$ & $1.001$ & $-0.271$ & $1.57$ \\
\tableline
\noalign{\smallskip}
\multicolumn{8}{c}{(b) Outlying SNe in spirals}\\
\noalign{\smallskip}
\tableline
1972E   & Am   & $ 0.00$ & $0.056$ & $-0.056$ & $1.048$ & $-0.383$ & $1.05$ \\
1990N   & $3$  & $ 0.02$ & $0.026$ & $-0.006$ & $1.195$ & $-0.254$ & $1.05$ \\
1990O   & $1$  & $ 0.08$ & $0.093$ & $-0.013$ & $1.117$ & $-0.411$ & $0.94$ \\
1990T   & $1$  & $-0.08$ & $0.053$ & $-0.133$ & $1.172$ & $-0.079$ & $1.13$ \\
1991S   & $3$  & $ 0.03$ & $0.026$ & $ 0.004$ & $1.154$ & $-0.274$ & $1.00$ \\
1991ag  & $3$  & $ 0.11$ & $0.062$ & $ 0.048$ & $1.109$ & $-0.351$ & $0.87$ \\
1992A   & $0$  & $ 0.01$ & $0.018$ & $-0.008$ & $1.043$ & $-0.273$ & $1.47$ \\
1992bc  & $2$  & $-0.12$ & $0.022$ & $-0.142$ & $0.990$ & $-0.339$ & $0.90$ \\
1992bl  & $0$  & $-0.03$ & $0.011$ & $-0.041$ & $1.036$ & $-0.234$ & $1.56$ \\
1995D   & $0$  & $ 0.04$ & $0.058$ & $-0.018$ & $1.146$ & $-0.375$ & $1.03$ \\
1998V   & $3$  & $ 0.17$ & $0.196$ & $-0.026$ & $1.088$ & $-0.165$ & $1.06$ \\
1998eg  & $5$  & $ 0.12$ & $0.123$ & $-0.003$ & \nodatr & $-0.150$ & $1.15$ \\
\tableline
\noalign{\smallskip}
\multicolumn{4}{r}{mean [$N=34,32,31$]:}     
                                   & $-0.013$ & $1.092$ & $-0.232$ & \\
\multicolumn{4}{r}{random error of mean:}
                                   &$\pm0.012$&$\pm0.010$&$\pm0.018$& \\
\multicolumn{4}{r}{$\sigma$:}                  
                                   &   $0.07$ &   $0.06$ &   $0.10$ &  \\ 
\enddata
  \def\baselinestretch{1.5}
\end{deluxetable}


     Since the SNe in Table~\ref{tab:color:minred} are assumed to
suffer no reddening in their host galaxies, their Galactic
absorption-corrected colors $(X\!-\!Y)^{0}$ are equal to the colors
$(X\!-\!Y)^{00}$ corrected for Galactic {\em and\/} host galaxy
reddening. The assumption of zero reddening is supported by several
arguments. 
(a) If one assumes that the average error of 
$B_{\max}$ and $V_{\max}$ is $0.05$, then the observed scatter of
$\sigma_{B\!-\!V}=0.07\mag$ in Table~\ref{tab:color:minred} is
fully explained by these errors. 
In that case the distribution of $(B\!-\!V)^{00}_{\max}$ in
Figure~\ref{fig:color:tab2}a must be Gaussian, which is 
statistically acceptable in spite of the four very blue SNe\,Ia with
$(B\!-\!V)^{00}_{\max} < -0.12$.
(b) As Figure~\ref{fig:color:tab2} shows, the color distribution of the
    SNeIa in Table~\ref{tab:color:minred} is skewed bluewards, if
    anything, while any noticeable reddening would skew the distribution
    towards red colors.  
(c) The colors $(B\!-\!V)_{\max}^{00}$ and $(V\!-\!I)_{\max}^{00}$
    of the SNe\,Ia in Table~\ref{tab:color:minred} are {\em
    uncorrelated} (in fact insignificantly anti-correlated); in case
    of absorption in the host galaxies both values would be redder
    than average.  
(d) If the spread in color in Figure~\ref{fig:color:tab2}a
    was due to absorption, the absolute magnitudes $M_{B}$ would
    correlate with the color $(B\!-\!V)^{0}_{\max}$ from column~5 with
    a slope of ${\cal R}_{B}=3.65$, whereas an insignificant slope of
    $0.137\pm0.904$ is observed.  

     The least squares solutions of the data in
Figure~\ref{fig:color:tab2dm15} are   
\begin{equation}\label{eq:color:mean:BV}
   (B\!-\!V)_{\max}^{00} = (0.045\pm0.044)\Delta m_{15} -
   (0.073\pm0.060),\; \sigma=0.07,\; N=34 
\end{equation}
\begin{equation}\label{eq:color:mean:BV35}
   (B\!-\!V)_{35}^{00} = (-0.068\pm0.036)\Delta m_{15} +
   (1.182\pm0.048),\; \sigma=0.05,\; N=32
\end{equation}
\begin{equation}\label{eq:color:mean:VI}
   (V\!-\!I)_{\max}^{00} = (0.150\pm0.062)\Delta m_{15} -
   (0.430\pm0.083),\; \sigma=0.09,\; N=31. 
\end{equation}
The dependence of $(B\!-\!V)_{\max}^{00}$ on $\Delta m_{15}$ in
equation~(\ref{eq:color:mean:BV}) is weaker than suggested by
\citet{Phillips:etal:99}, \citet{Nobili:etal:03}, and
\citet{Altavilla:etal:04}. In fact the dependence is only marginally
significant. Yet it is adopted here at face value. -- The similarity
of SN\,Ia spectra at $t_{B}=35^{\rm d}$ and the similar slopes of the
color curves at this late phase have led \citet{Lira:95},
\citet{Riess:etal:96}, \citet{Phillips:etal:99}, and
\citet{Altavilla:etal:04} to assume that $(B\!-\!V)^{00}_{35}$ is
independent of $\Delta m_{15}$. However,
equation~(\ref{eq:color:mean:BV35}) shows this color to become {\em
  bluer\/} with increasing $\Delta m_{15}$ at a significance of
$2\sigma$. -- The $(V\!-\!I)_{\max}^{00}$ colors exhibit in
equation~(\ref{eq:color:mean:VI}) a clear dependence on $\Delta
m_{15}$, becoming {\em redder\/} with increasing decline rate.
 
The mean intrinsic colors of normal SNe\,Ia, normalized to $\Delta
m_{15}=1.1$, become from equations
(\ref{eq:color:mean:BV}$-$\ref{eq:color:mean:VI}) 
\begin{equation}\label{eq:color:mean}
   (B\!-\!V)_{\max}^{00} = -0.024\pm0.010, 
   (B\!-\!V)_{35}^{00} = 1.107\pm0.009, \mbox{and }
   (V\!-\!I)_{\max}^{00} = -0.265\pm0.016.
\end{equation}

     Having argued that SNe\,Ia in Table~\ref{tab:color:minred} are
essentially redding-free, it is nevertheless clear that their
reddening is not exactly zero. Their true colors must therefore
be slightly bluer than expressed in
equation~(\ref{eq:color:mean}). We will return to this point in
\S~\ref{sec:conclusions} where it is stressed that any rest reddening
is inconsequential for the use of SNe\,Ia as standard candles as long
as they are reduced to some {\em uniform\/} color.

\section{THE REDDENING OF SNe\,Ia IN THEIR HOST GALAXIES}
\label{sec:hostgal}
The reddening $E(B\!-\!V)_{\rm host}$ of SNe\,Ia at maximum phase due
to selective absorption in the host galaxy is determined in three
different ways.

\subsection{The Reddening in the Host Galaxy from \boldmath{$(B\!-\!V)^{00}_{\max}$}}
\label{sec:hostgal:max}
The reddening in the host galaxy is given by
\begin{equation}\label{eq:EBVmax}
   E(B\!-\!V)_{\max}= B_{\max} - V_{\max} - E(B\!-\!V)_{\rm Gal} - (B\!-\!V)_{\max}^{00}. 
\end{equation}
$E(B\!-\!V)_{\max}$ follows for all SNe\,Ia in Table~\ref{tab:SN1}a
and tentatively for the peculiar SNe\,Ia in Table~\ref{tab:SN1}b by
inserting the apparent magnitudes and the Galactic reddening from
columns~6, 7, and 10 in Table~\ref{tab:SN1}a,b and by taking the 
intrinsic color from equation~(\ref{eq:color:mean:BV}). The resulting
$E(B\!-\!V)_{\max}$ are given in column~11 of Table~\ref{tab:SN1}a \& b.

\subsection{The Reddening in the Host Galaxy from the Tail Colors \boldmath{$(B\!-\!V)^{00}_{35}$}} 
\label{sec:hostgal:BVmax}
Color excesses $E(B\!-\!V)_{35}$ at phase $t_{B}=35\;$days were kindly
provided by S.~Jha for 59 SNe\,Ia in Table~\ref{tab:SN1}a. He has
based them on an adopted intrinsic tail color of
$(B\!-\!V)^{00}_{35}=1.055$ for all SNe\,Ia. Additional tail 
color excesses were published by \citet{Phillips:etal:99} and
\citet{Altavilla:etal:04}. They are in a slightly different system as
those by Jha. In a first step the Altavilla excesses were transformed
into Jha's system by the regression
\begin{equation}\label{eq:color:alta}
   \Delta E(B\!-\!V)_{{\rm Alta}-{\rm Jha}} =
   (0.048\pm0.032)E(B\!-\!V)_{\rm Alta} + 
   (0.020\pm0.006), \;  \sigma=0.03,\; N=38.
\end{equation}
In a second step the Phillips excesses were transformed by comparison
with the joint set from Jha and Altavilla by means of 
\begin{equation}\label{eq:color:phil}
   \Delta E(B\!-\!V)_{{\rm Phil}-{\rm Jha}} =
   (0.084\pm0.025)E(B\!-\!V)_{\rm Phil} + 
   (0.018\pm0.009), \;  \sigma=0.06,\; N=62.
\end{equation}
The homogenized excesses $E(B\!-\!V)_{35}$ of SNe\,Ia with more than
one determination were averaged. If one adds to these values Jha's
adopted intrinsic color of
$(B\!-\!V)^{00}_{35}=1.055$, one recovers the ``observed'' tail colors
$(B\!-\!V)^{0}_{35}$, not listed, for 105 of the SNe\,Ia.

     New color excesses $E(B\!-\!V)_{35}$ are now determined by
making allowance for the $\Delta m_{15}$ dependence of the intrinsic
color $(B\!-\!V)^{00}_{35}$. This is achieved by subtracting
$(B\!-\!V)^{00}_{35}$ as given in equation~(\ref{eq:color:mean:BV35})
from the ``observed'' colors $(B\!-\!V)^{0}_{35}$. 

\begin{figure}[t]
     \epsscale{0.55}
     \plotone{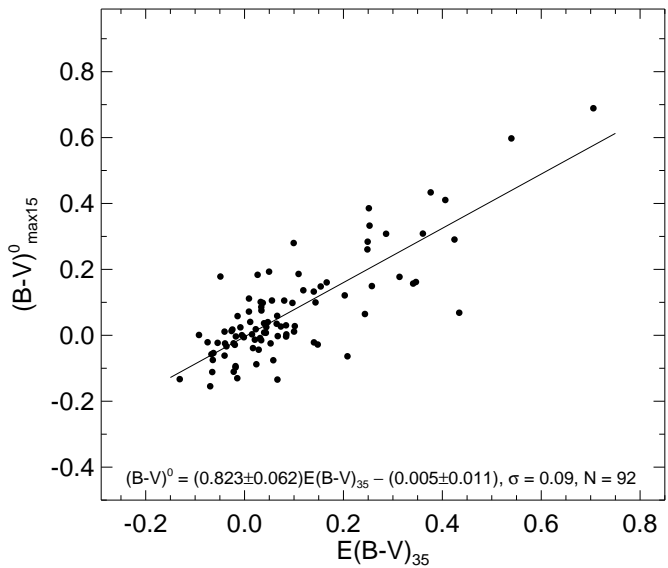}
     \caption{The relation of the colors $(B\!-\!V)_{\max15}^{0}$
     versus the reddening values $E(B\!-\!V)_{35}$ which are derived
     from a {\em red\/} SN phase.}   
\label{fig:color:E35}
\end{figure}

\begin{figure}[t]
     \plotone{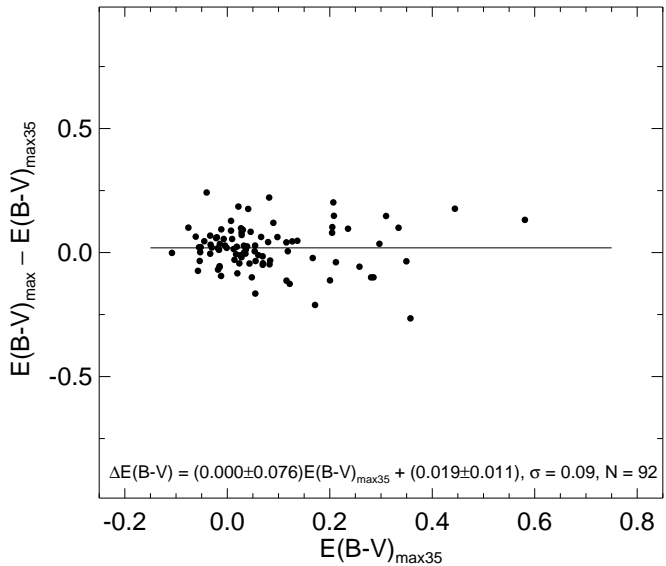}
     \caption{Comparison of the color excesses  $E(B\!-\!V)_{\max}$
     from colors at maximum and  $E(B\!-\!V)_{max35}$ from tail colors
     at $t_{B}=35^{\rm d}$.}   
\label{fig:color:comp1}
\end{figure}

     The adopted tail excesses $E(B\!-\!V)_{35}$ apply to
the quite red phase at $t_{B}=35^{\rm d}$. They cannot directly be
applied to the (blue) SN colors at maximum, because $E(B\!-\!V)$ must
depend on color due to bandwidth effects. In fact, if one plots the
colors $(B\!-\!V)_{\max15}^{0}$ (corrected for Galactic reddening and
reduced to $\Delta m_{15}=1.1$ with equation~(\ref{eq:color:mean:BV}))
against the adopted values of $E(B\!-\!V)_{35}$ one obtains the
following linear regression (Figure~\ref{fig:color:E35}).    
\begin{equation}
 \label{eq:color:BVmax}
   (B\!-\!V)_{\max15}^{0} = (0.823\!\pm\!0.062)E(B\!-\!V)_{35} -
   (0.005\!\pm\!0.011), \;   \sigma=0.09, \; N=92
\end{equation}
Since the {\em intrinsic\/} color $(B\!-\!V)^{00}$ at maximum must 
not be a function of reddening it follows from
equation~(\ref{eq:color:BVmax}) that
\begin{equation}
 \label{eq:color:EBVmax35}
E(B\!-\!V)_{\max35}  =  (0.823\pm0.062)E(B\!-\!V)_{35}.
\end{equation}
The resulting values $E(B\!-\!V)_{\max35}$ are listed in
Table~\ref{tab:SN1}a, column~12. 
The excesses from equations~(\ref{eq:EBVmax} \&
\ref{eq:color:EBVmax35}) are compared in
Figure~\ref{fig:color:comp1}. The agreement is satisfactory with no
systematic differences.

     The excesses $E(B\!-\!V)_{\max35}$ for four peculiar SNe\,Ia in
Table~\ref{tab:SN1}b are given in parentheses. They are smaller on
average than their excesses from $(B\!-\!V)_{\max}$, suggesting that
they are redder at $t_{B}=35^{\rm d}$ than normal SNe\,Ia (see also
\S~\ref{sec:SNpec:T}). The excesses $E(B\!-\!V)_{\max35}$ of the
SN\,1991bg-like SNe in Table~\ref{tab:SN1}c have not been used, but
they are consistent with the assumption that these SNe in early-type
galaxies are almost reddening-free.

\subsection{The Reddening in the Host Galaxy from 
  \boldmath{$(V\!-\!I)^{00}_{\max}$}} 
\label{sec:hostgal:EVImax}
A third possibility to determine the reddening $E(B\!-\!V)$
in the host galaxy is provided by the magnitudes $V_{\max}$ and
$I_{\max}$ in Table~\ref{tab:SN1}a\,\&\,b, columns 7 \& 8. We note 
\begin{equation}\label{eq:color:EVIhost}
   E(V\!-\!I) = V_{\max} - I_{\max} - 1.3E(B\!-\!V)_{\rm
   Gal} - (V\!-\!I)_{\max}^{00},  
\end{equation}
where $E(V\!-\!I)=1.3E(B\!-\!V)$ is adopted for the Galactic
reddening. If one inserts the intrinsic colors from
equation~(\ref{eq:color:mean:VI}), with the $\Delta m_{15}$-values
appropriate for each SN\,Ia, into equation~(\ref{eq:color:EVIhost}) one
obtains $E(V\!-\!I)$.

\begin{figure}[t]
     \plotone{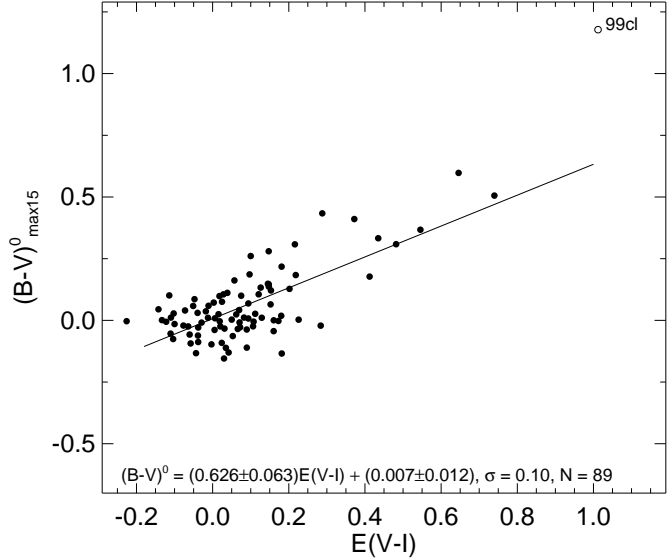}
     \caption{The correlation of the color $(B\!-\!V)_{\max15}^{0}$ and
     $E(V\!-\!I)$. SN\,1999cl is not taken into account.}  
\label{fig:color:EVIhost}
\end{figure}

\begin{figure}[t]
     \plotone{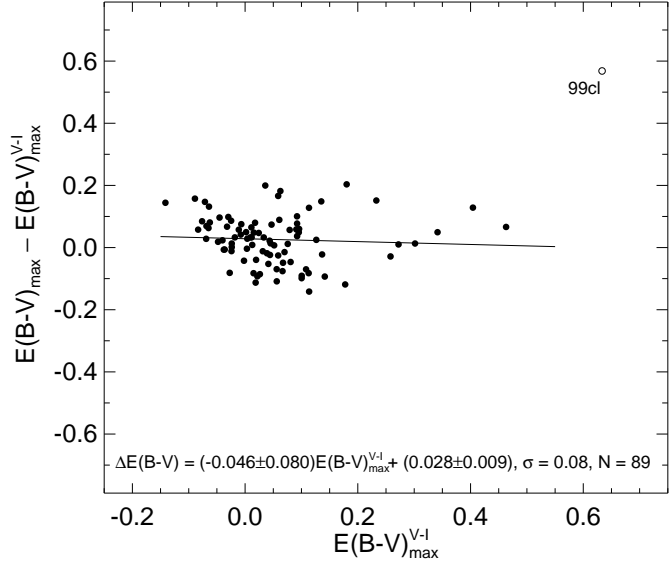}
     \caption{The difference $E(B\!-\!V)_{\max}$  minus
     $E(B\!-\!V)^{V\!-\!I}_{\max}$ plotted against 
     $E(B\!-\!V)^{V\!-\!I}_{\max}$. $E(B\!-\!V)_{\max}$ is here the
     mean of $E(B\!-\!V)_{\max}$ and $E(B\!-\!V)_{\max35}$.}  
\label{fig:color:comp2}
\end{figure}

     Since $E(B\!-\!V)$ is required, not $E(V\!-\!I)$, the latter must
be converted into the former. This is done by means of
Figure~\ref{fig:color:EVIhost}, where 
$(B\!-\!V)^{0}_{\max15}$ (as in Figure~\ref{fig:color:E35}) is
plotted against $E(V\!-\!I)$. 
From the equation at the
bottom of Figure~\ref{fig:color:EVIhost} follows
\begin{equation}\label{eq:color:EBVVIhost}
   E(B\!-\!V)^{V\!-\!I}_{\max} = (0.626\pm0.063)E(V\!-\!I).  
\end{equation}
The color excesses  $E(B\!-\!V)^{V\!-\!I}_{\max}$, listed in column~13
of Table~\ref{tab:SN1}, are compared with the mean of the excesses
$E(B\!-\!V)_{\max}$  and $E(B\!-\!V)_{\max35}$ in
Figure~\ref{fig:color:comp2}. Again the agreement is satisfactory.

\subsection{The Adopted Reddening in the Host Galaxies} 
\label{sec:hostgal:adopted}
In view of the consistency of the color excesses $E(B\!-\!V)_{\max}$,
$E(B\!-\!V)_{\max35}$, and $E(B\!-\!V)^{V\!-\!I}_{\max}$ they are
averaged with equal weights. 
The means are adopted as the best estimates of the reddenings in the
host galaxy, $E(B\!-\!V)_{\rm host}$.
The values of $E(B\!-\!V)_{\rm host}$ are listed in
Table~\ref{tab:SN3}, column~2. The ensuing colors
$(B\!-\!V)^{00}_{\max}$, fully corrected
for Galactic and host galaxy reddening, are in column~3.

Table~\ref{tab:SN3} is organized as follows:\\
Col.~1: The name of the SN.\\
Col.~2: The adopted reddening $E(B\!-\!V)_{\rm host}$ in the host
        galaxy being the mean of columns 11-13 of
        Table~\ref{tab:SN1}.\\ 
Col.~3\&4: The intrinsic colors $(B\!-\!V)^{00}_{\max}$ and
        $(V\!-\!I)^{00}_{\max}$ corrected for Galactic reddening and
        reddening in the host galaxy.\\
Col.~5-7: The apparent magnitudes $m^{00}_{BVI}$ at maximum, corrected
        for absorption in the Galaxy and in the host galaxy.\\
Col.~8-10: The absorption-corrected apparent magnitudes 
        $m^{\rm corr}_{BVI}$ reduced to a standard decline rate of
        $\Delta m_{15}=1.1$ and $(B\!-\!V)^{00}_{\max}=-0.024$ by
        the precepts developed in \S~\ref{sec:other:corrected}.\\
Col.~11: The photometric absorption-free distance modulus
        $(m-M)^{00}_{\rm lum}$ from $m^{\rm corr}_{V}$ in column~9 and 
        the mean luminosity $M^{\rm corr}_{V}$ in
        equation~(\ref{eq:other:Mabs}). The latter assumes $H_{0}=60$.

\begin{figure}[t]
     \plotone{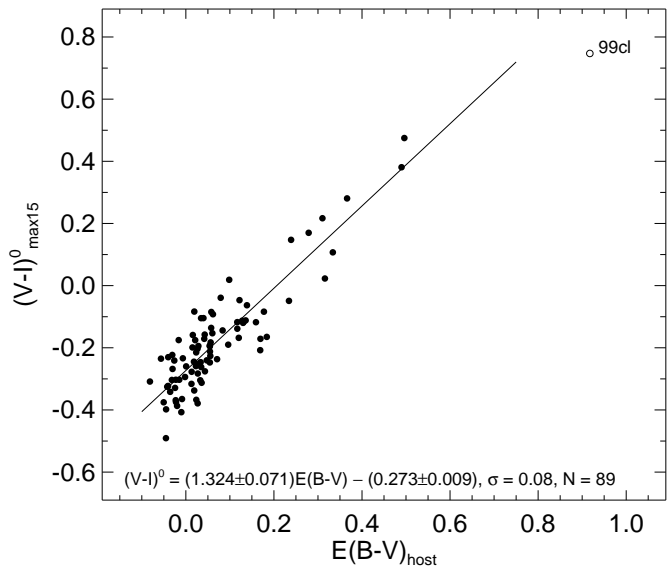}
     \caption{The colors $(V\!-\!I)^0_{\max15}$, corrected for Galactic
     reddening, versus the adopted reddening $E(B\!-\!V)_{\rm host}$
     in the host galaxy.}   
\label{fig:color:EVIBVhost}
\end{figure}
     In order to free the $(V\!-\!I)^{0}$ colors from the host galaxy
reddening the relation between $E(V\!-\!I)_{\rm host}$ and
$E(B\!-\!V)_{\rm host}$ is needed. It is obtained from
Figure~\ref{fig:color:EVIBVhost}, where $(V\!-\!I)^{0}$ is plotted
versus $E(B\!-\!V)_{\rm host}$. The figure yields (slightly rounded) 
\begin{equation}\label{eq:color:EVIBVhost}
   E(V\!-\!I)_{\rm host} = (1.32\pm0.07)E(B\!-\!V)_{\rm host},  
\end{equation}
which we adopt. The conversion factor is consistent with
equation~(\ref{eq:color:EBVVIhost}), but here $E(B\!-\!V)_{\rm host}$
has high weight being the mean of three determinations. The rms
dispersion of $E(B\!-\!V)_{\rm host}$, if based on the three
determinations,  is $0.043\;$mag. 

\subsection{Test of the K-Correction} 
\label{sec:color:Kcorr}
\begin{figure*}[t]
     \epsscale{1.0}
     \plotone{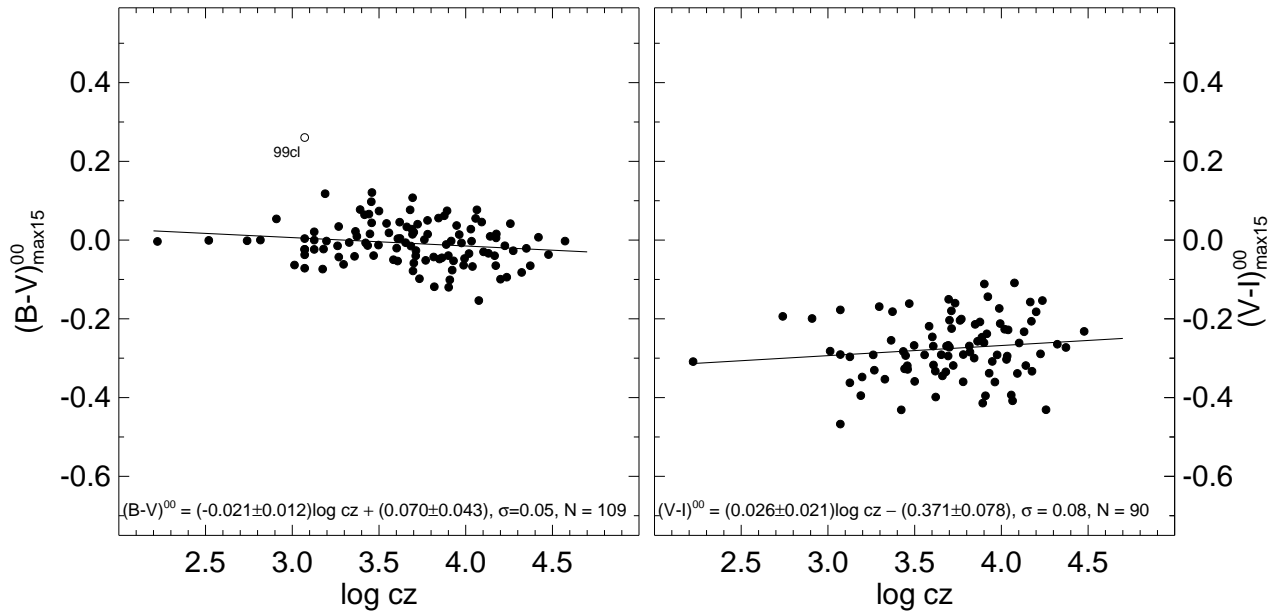}
     \caption{A test of the adopted K-corrections. The colors
     $(B\!-\!V)^{00}_{\max15}$ and $(V\!-\!I)^{00}_{\max15}$,
     normalized to $\Delta m_{15}=1.1$, show only a negligible 
     correlation.}  
\label{fig:color:Kcorr}
\end{figure*}

The template-fitting of the SN light curves by the method of
\citet{Hamuy:etal:96c} is applied in a way as to
compensate the K-correction, i.e.\ the effect of the redshifts $z$ on
the magnitudes and colors, in a single step. 
To verify the success of the method we plot the adopted
colors $(B\!-\!V)_{\max15}^{00}$ and  $(V\!-\!I)_{\max15}^{00}$,
normalized to $\Delta m_{15}=1.1$, against
$\log cz$ in Figure~\ref{fig:color:Kcorr}. Any dependence is hardly
significant and is neglected in the following.

\section{THE ABSORPTION-CORRECTED MAGNITUDES OF SNe\,Ia}
\label{sec:absorption}
%
\subsection{The Galactic Absorption}
\label{sec:absorption:galactic}
The Galactic reddenings of \citet{Schlegel:etal:98} are converted into 
absorption values $A_{B}$, $A_{V}$, and $A_{I}$ by means of
conventional absorption-reddening ratios of ${\cal R}_{B} = 4.1$,
${\cal R}_{V} = 3.1$, and ${\cal R}_{I} = 1.8$ (hence $E(V\!-\!I) =
1.3E(B\!-\!V)$ as used in \S~\ref{sec:color}). The resulting
absorption values are subtracted from the observed SN magnitudes to
obtain $B^{0}$, $V^{0}$, and $I^{0}$ at maximum.  

     Conventional values of ${\cal R}$ have been adopted because the
available range of $E(B\!-\!V)_{\rm Gal}$ is too small to derive
meaningful Galactic ${\cal R}$ values from the SNe\,Ia themselves.  

\subsection{The Absorption in the Host Galaxy}
\label{sec:absorption:host}
Previous attempts to correct SN magnitudes for absorption in the host
galaxies have frequently assumed conventional (Galactic) values of
${\cal R}_{BVI}$, although 
\citet[][and references therein]{Branch:Tammann:92} have argued for
significantly smaller values. More recent determinations are (quoted
${\cal R}_{V}$ values are transformed here to ${\cal R}_{B}$ by 
${\cal R}_{B}\equiv {\cal R}_{V}+1$)
${\cal R}_{B}=3.55\pm0.30$ \citep{Riess:etal:96}, 
$2.09$ \citep{Tripp:98},
$3.5\pm0.4$ \citep{Phillips:etal:99}, 
$2.8$ \citep{Krisciunas:etal:00},
$3.88\pm0.15$ \citep{Wang:etal:03},
$3.5$ \citep{Altavilla:etal:04}. 
The present large sample of SNe\,Ia is well suited for an independent
determination of the average value of ${\cal R}_{BVI}$ of
extragalactic SNe\,Ia.

\begin{figure*}[t]
     \epsscale{1.0}
     \plotone{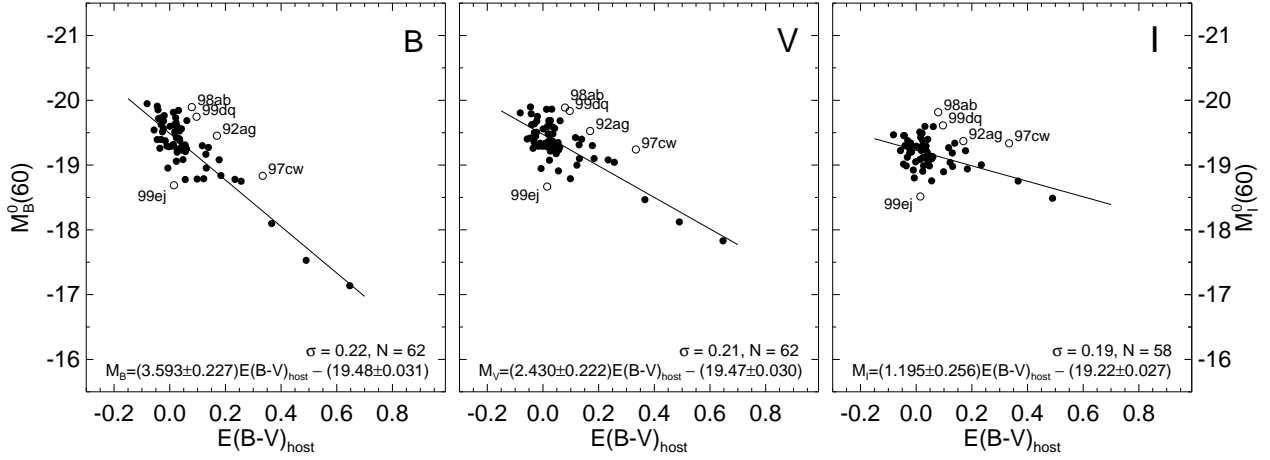}
     \caption{The dependence of the absolute magnitudes
     $M^{0}_{BVI}(60)$ on the host galaxy reddening
     $E(B\!-\!V)_{\rm host}$ for SNe\,Ia with $3000 < v_{\rm
     CMB}<20\,000\kms$. The $M^{0}_{BVI}$ magnitudes are only
     corrected for Galactic absorption. Five SNe\,Ia shown here as
     open symbols are not used for the solution (see text).}
\label{fig:absorption:Mabs}
\end{figure*}

     For this purpose the absolute magnitudes $M^{0}_{BVI}$, based on
$H_{0}=60$ and corrected only for Galactic absorption, are plotted
against the values of the host galaxy reddening $E(B\!-\!V)_{\rm
  host}$ as listed in column~13 of Table~\ref{tab:SN1}
(Figure~\ref{fig:absorption:Mabs}).  

   The SN sample used here is called in the following the ``fiducial''
sample. It contains the 62\,(58) SNe\,Ia in Table~\ref{tab:SN1}a with
$B$, $V$ (and $I$) magnitudes and with $3000<v_{\rm
  CMB}<20\,000\kms$. It excludes SN\,1996ai which suffers excessive
absorption in its host galaxy (see Table~\ref{tab:SN1}a). It also
excludes five SNe\,Ia, which deviate by more 
than $2\sigma_{M}$. The apparently normal SN\,1999ej is
too faint for unknown reasons by $3\sigma_{M}$. SN\,1992ag is very red
at $t_{35}$, causing its tail excess $E(B\!-\!V)_{max35}$ to be large,
which in turn makes $(B\!-\!V)_{\max}^{00}$ exceptionally blue
and the absolute magnitude very bright (cf.\ Table~\ref{tab:SN1}a and
\ref{tab:SN3}). It is believed that either its color
$(B\!-\!V)_{35}^{0}$ is in error or its light curve is peculiar. If
the color excess was based only on the color at maximum its absolute
magnitude would become normal. The overly bright SN\,1997cw is
suspicious not so much because its photometry begins only 16 days
after $B$ maximum, but because it was classified as SNe-91T
\citep{Berlind:etal:97}, which is now revised to SNe-91T/99aa
\citep{Li:etal:01b}. Also the very luminous SN\,1998ab and 1999dq are
of type SNe-99aa \citep{Li:etal:01b}.  The properties of this new
class of SNe\,Ia and their overluminosity are taken up again in
\S~\ref{sec:SNpec:T}. 

     Four other less overluminous SNe-99aa, designated with a dagger
$\dagger$ in Table~\ref{tab:SN1}a, are kept in the sample of normal
SNe\,Ia. The reason is that they can be isolated only from pre-maximum
spectroscopy, which is lacking for many objects in
Table~\ref{tab:SN1}a. Thus a few of the ``normal'' SNe\,Ia may
actually belong to the elusive class of SNe-99aa without having as
extreme luminosities like SN\,1997cw, 1998ab, and 1999dq.

     As shown by the equations in
Figure~\ref{fig:absorption:Mabs} the resulting values of ${\cal
  R}_{BVI}$ are about 3.6, 2.4, and 1.2. However, these values are
still open to criticism. As will be shown below SNe\,Ia in E/S0 have
lower intrinsic luminosities {\em and\/} less extinction than
average. Therefore there is some dependence of $M^{0}_{BVI}$ on
$E(B\!-\!V)$ which cannot be attributed to absorption, but which is
due to the intrinsic underluminosity of SNe\,Ia in early-type
galaxies. This type-dependent effect can be compensated by the decline
rate $\Delta m_{15}$ in \S~\ref{sec:other:lightcurve}. The interplay
of ${\cal R}$ and  $\Delta m_{15}$ requires therefore a simultaneous
solution which is given in equation (\ref{eq:other:dm15R}) below. From
this the improved values of ${\cal R}_{BVI}$ become  
\begin{equation}\label{eq:absorption:R}
   {\cal R}_{B}=3.65\pm0.16,\quad 
   {\cal R}_{V}=2.65\pm0.15,\quad 
   {\cal R}_{I}=1.35\pm0.21. 
\end{equation}
The ${\cal R}$ values here are in addition adjusted to fulfill the
condition ${\cal R}_{B}-{\cal R}_{V}=1$. They are quite close to the
provisional values in Figure~\ref{fig:absorption:Mabs}. It is also
satisfactory that they agree quite well with those by other authors
cited above.

     If one assumes that the highly reddened SN\,1996ai is a
normal SNe\,Ia with average luminosity and that its 
peculiar motion is $\la200$\kms, one finds ${\cal R}_{B}=3.0\pm0.4$
from this single object.

     The solutions in Table~\ref{tab:other:dm15R} for ${\cal R}_{V}$
and ${\cal R}_{I}$ imply a ratio $E(V\!-\!I)/E(B\!-\!V)={\cal
  R}_{V}-{\cal R}_{I}=1.21\pm0.26$, which may be compared with
$1.32\pm0.07$ from equation~(\ref{eq:color:EVIBVhost}). The small
error of the latter value is due to the fact that its determination is
independent of any assumed value of ${\cal R}$, which is notoriously
difficult to determine. It is noted that the adopted value of $1.30$
is the same as the conventional Galactic value of 1.3.

     The values of $E(B\!-\!V)_{\rm host}$ (Table~\ref{tab:SN3},
column~2) were multiplied with the ${\cal R}_{BVI}$ values in
equation~(\ref{eq:absorption:R}) to be subtracted from the apparent
magnitudes $B$, $V$, and $I$  (Table~\ref{tab:SN1}, columns~6-8), 
after they are corrected for the conventional
Galactic absorption. The resulting fully corrected apparent magnitudes
$m_{B\max}^{00}$, $m_{V\max}^{00}$, and $m_{I\max}^{00}$ are tabulated in
columns~5-7 of Table~\ref{tab:SN3}.

\section{THE DEPENDENCE OF SN LUMINOSITIES ON OTHER PARAMETERS}
\label{sec:other}
%
\subsection{The Dependence of SN Luminosities on Galaxy Type}
\label{sec:other:galaxyT}
\begin{figure}[t]
     \epsscale{0.55}
     \plotone{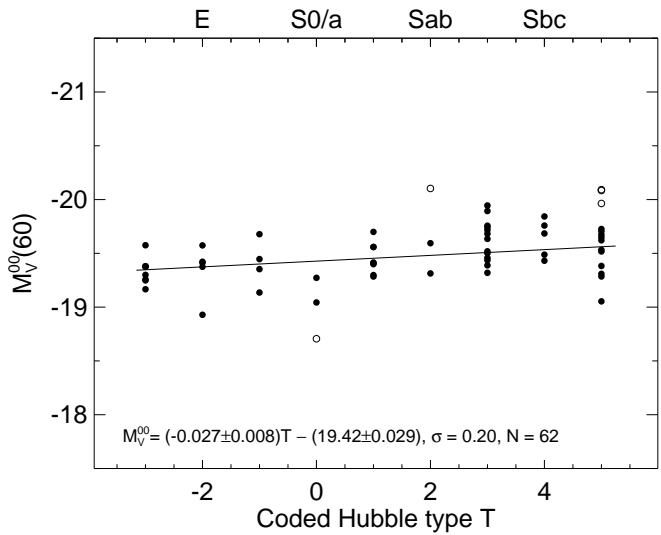}
     \caption{The correlation between the absolute magnitudes
     $M_{V}^{00}(60)$, fully corrected for absorption, and the
     coded Hubble type of the host galaxy.} 
\label{fig:other:type}
\end{figure}

If one plots the absolute magnitudes $M_{V}^{00}$ as
introduced in \S~\ref{sec:absorption}, but now also corrected for
absorption in the host galaxy and {\em without\/} the $\Delta m_{15}$
correction, against the Hubble type of the host galaxy, a clear
correlation emerges between SN luminosity  and the Hubble type as
first shown by \citet{Hamuy:etal:96a,Hamuy:etal:00}, the SNe\,Ia in
early-type galaxies being fainter by $\sim0.3\mag$ than in late spirals
(Figure~\ref{fig:other:type}). The obvious conclusion is that young
populations produce brighter SNe\,Ia. 

     Yes, but why? The most secure conclusion from the data, 
which are beyond doubt 
\citep[][Fig. 6]{Hamuy:etal:95,Saha:etal:97,Saha:etal:99,Parodi:etal:00, Sandage:etal:01}, 
is that different channels of SNe\,Ia production exist between 
most SNe\,Ia in E/S0 galaxies compared with most SNe Ia in spirals.

     Both channels are fed by old, highly evolved stars near one 
solar mass that are in close binaries. By one of two processes 
(mass transfer from a normal star to a white dwarf, or mergers of 
two white dwarfs due to orbital energy loss caused by 
gravitational radiation nudging one or both stars over the 
Chandrasekhar limit), catastrophic collapse eventually occurs, 
but with different gestation times. Hence, the delay times differ 
for progenitor formation to explosion, according to the process. 
The consequence is that the star formation rates of the initial 
progenitors differ between E/S0 galaxies and spirals (E/S0 with a 
single burst of star formation early on; spirals with a more  
continuous star formation rate). Hence, the average delay times  
differ between E/S0 and spirals (L. Greggio, private communication),
making a difference in the Fe production rate, hence a chemical 
composition difference. This expected chemical composition 
difference apparently affects the absolute magnitude at maximum 
in the SN explosion, as suggested by \citet{Hamuy:etal:00}.   

\subsection{The Dependence of SN Luminosity on Light Curve Parameters}
\label{sec:other:lightcurve}
\begin{figure*}[t]
     \epsscale{1.0}
     \plotone{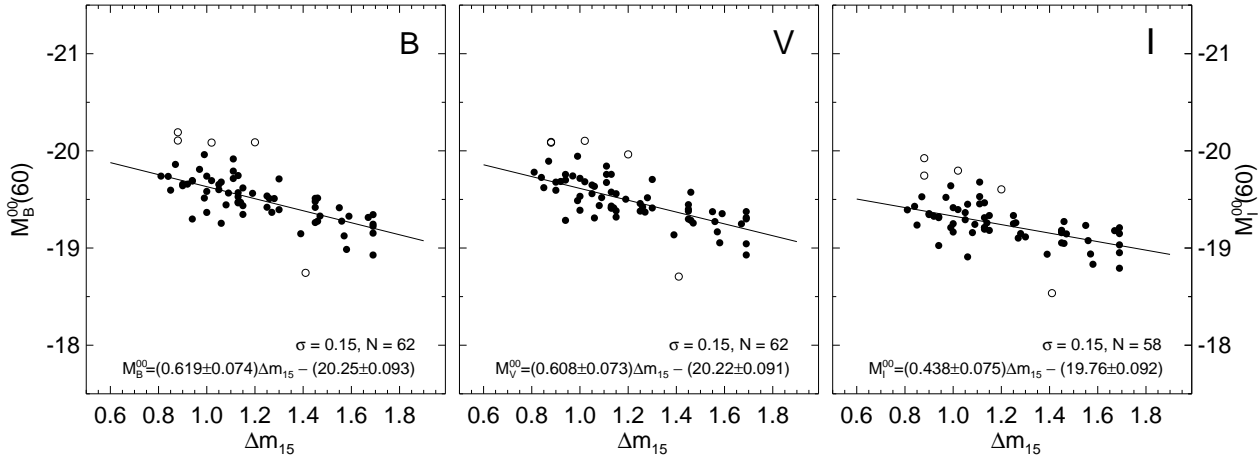}
     \caption{The dependence of the  absolute magnitudes
     $M_{BVI}^{00}(60)$ on the decline rate $\Delta m_{15}$. The
     SNe\,Ia with open symbols have not been used for the
     solution. They are identified in Figure~\ref{fig:absorption:Mabs}
     and discussed in \S~\ref{sec:absorption:host}.}  
\label{fig:other:dm15}
\end{figure*}

Since also the decline rate is a function of the Hubble type, one
expects a correlation between the luminosity and the decline rate as
well. This Pskovskii effect was first quantified by the
decline rate parameter $\Delta m_{15}$ by \citet{Phillips:93}. 
The corresponding correlation of the form
\begin{equation}\label{eq:absorption:dm15}
   M^{00}_{BVI15}=M^{00}_{BVI} - \alpha_{BVI}(\Delta m_{15}-1.1)
\end{equation}
is in fact even tighter than the one in
Figure~\ref{fig:other:type}; it is shown in Figure~\ref{fig:other:dm15}.
The normalization to $\Delta  m_{15}=1.1$ corresponds roughly to the
median value of normal SNe\,Ia. The magnitudes $M^{00}_{BVI}$ are
corrected for Galactic and internal absorption described in
\S~\ref{sec:absorption}. 

     Since $\alpha$ and ${\cal R}$ and hence $M^{00}$ are not quite
independent, as discussed in \S~\ref{sec:absorption:host}, these
quantities must be simultaneously solved for. The corresponding
regression with two independent variables takes the form of 
\begin{equation}\label{eq:other:dm15R}
   M^{0}_{BVI}= \alpha_{BVI}(\Delta m_{15}-1.1) +
                {\cal R}_{BVI}E(B\!-\!V)_{\rm host} +
                M^{00}_{BVI15},
\end{equation}
where the constant term, $M^{00}_{BVI15}$, is the mean absolute
magnitude of SNe\,Ia, corrected for Galactic and host galaxy
absorption and reduced to $\Delta m_{15}=1.1$. Least-squares solutions
for the same fiducial SN sample as used for Figure~\ref{fig:absorption:Mabs}
give the values of $\alpha_{BVI}$, ${\cal R}_{BVI}$, and
$M^{00}_{BVI15}$ as shown in Table~\ref{tab:other:dm15R}.
\def\baselinestretch{1.1}
\begin{deluxetable}{lccc}
\tablewidth{0pt}
\tabletypesize{\footnotesize}
\tablenum{4}
\tablecaption{The coefficient $\alpha$ of the decline rate, the
  absorption-to-reddening ratio ${\cal R}$, and the
  absorption-corrected mean absolute magnitude of normal SNe\,Ia with
  $\Delta m_{15}=1.1$.\label{tab:other:dm15R}} 
\tablehead{
 & \colhead{$B$} & \colhead{$V$} & \colhead{$I$} 
}
\startdata
$\alpha$          & $0.619\pm0.076$ & $0.608\pm0.074$ & $0.438\pm0.076$ \\
${\cal R}$        &  $3.75\pm0.16$  &  $2.59\pm0.15$  &  $1.38\pm0.21$ \\
$M^{00}_{15}(60)$ &$-19.57\pm0.02$  &$-19.55\pm0.02$  &$-19.28\pm0.02$ \\
$\sigma_{M}$      &      $0.15$     &      $0.15$     &     $0.15$     \\
$N$               &       62        &      62         &      58        \\
\enddata
  \def\baselinestretch{1.5}
\end{deluxetable}

     The values of $\alpha_{BVI}$ in Table~\ref{tab:other:dm15R} will
be adopted in the following. The dependence of
$M^{00}_{BVI}=M^{0}_{BVI}-{\cal R}_{BVI}E(B\!-\!V)_{\rm host}$ on $\Delta
m_{15}$ is illustrated in Figure~\ref{fig:other:dm15}, where
$\alpha_{BVI}$ appears as the slope of the correlation.

     The values of ${\cal R}_{BVI}$ in Table~\ref{tab:other:dm15R}
differ slightly from the values in equation~(\ref{eq:absorption:R}),
which we have adopted as the final ones, because the latter comply
with the additional conditions that ${\cal R}_{B}-{\cal R}_{V}=1$ and
${\cal R}_{V}-{\cal R}_{I}=1.3$ from
equation~(\ref{eq:color:EVIBVhost}).  

     In equation~(\ref{eq:color:mean:BV}) it was shown that the color
$(B\!-\!V)^{00}_{\max}$ is only marginally dependent on $\Delta
m_{15}$. This is vindicated here by the $\alpha$'s in
Table~\ref{tab:other:dm15R}. They yield for the relation
\begin{equation}\label{eq:absorption:BVbeta}
   (B\!-\!V)^{00}_{\max} = \beta\Delta m_{15} + \mbox{const}
\end{equation}
an insignificant value of
$\beta=\alpha_{B}-\alpha_{V}=0.011\pm0.106$. The corresponding
coefficient for the $(V\!-\!I)^{00}_{\max}$ colors becomes 
$\alpha_{V}-\alpha_{I}=0.170\pm0.106$, which is consistent with
$0.150\pm0.062$ from equation~(\ref{eq:color:mean:VI}).

     It becomes clear from the above that any change of $\beta$
affects necessarily the coefficients $\alpha$, because an
excess $\Delta\beta$ causes a change of the color excess of 
\begin{equation}\label{eq:other:beta1}
   \Delta E(B\!-\!V)_{\max} \propto -\Delta\beta\Delta m_{15}.
\end{equation}
This affects the absorption-corrected magnitudes $M_{B}^{00}$, taken
as an example, by 
\begin{equation}\label{eq:other:beta2}
   \Delta M_{B}^{00} \propto {\cal R}_{B}\Delta\beta\Delta m_{15}.
\end{equation}
Hence, if the {\em intrinsic\/} dependence of $M_{B}$ on $\Delta
m_{15}$ is $\Delta M_{B}^{00} \propto \alpha\Delta m_{15}$, one
observes instead
\begin{equation}\label{eq:other:beta3}
   \Delta M_{B}^{00} \propto (\alpha_{B} + {\cal R}_{B}\Delta\beta)\Delta m_{15}.
\end{equation}
The analogue holds for the color excesses $E(B\!-\!V)_{35}$ derived
from tail colors $(B\!-\!V)_{35}^{0}$. If the true colors vary like
$\Delta(B\!-\!V)_{35}^{0}\propto \gamma\Delta m_{15}$, any deviation
$\Delta \gamma$ from $\gamma$ leads to a variation of the color excess
of 
\begin{equation}\label{eq:other:beta4}
   \Delta E(B\!-\!V)_{35} \propto -\Delta\gamma\Delta m_{15},
\end{equation}
such that $\Delta E(B\!-\!V)_{\rm host}$, being the mean of 
$\Delta E(B\!-\!V)_{\max}$ and $\Delta E(B\!-\!V)_{35}$, varies like
\begin{equation}\label{eq:other:beta5}
   \Delta E(B\!-\!V)_{\rm host} =
   -\frac{1}{2}(\Delta\beta+\Delta\gamma)\Delta m_{15}, \quad \mbox{and}
\end{equation}
\begin{equation}\label{eq:other:beta6}
   \Delta M_{B}^{00} \propto [\alpha_{B} + \frac{1}{2}{\cal
   R}_{B}(\Delta\beta+\Delta\gamma)]\Delta m_{15} = (\alpha_{B} +
   \Delta\alpha)\Delta m_{15}.
\end{equation}
We have adopted, besides ${\cal R}_{B}=3.65$, $\beta=0.045$
(equation~\ref{eq:color:mean:BV}),
$\gamma=-0.068$ (equation~\ref{eq:color:mean:BV35}) 
as the intrinsic values. Yet some authors have adopted
other values of ${\cal R}$, $\beta$ and $\gamma$ which necessarily
reflect on the observed value of $\alpha$. 
For instance \cite{Phillips:etal:99} have
chosen $\Delta\beta=0.069$, $\Delta\gamma=0.068$, and ${\cal
  R}_{B}=4.16$. If one subtracts the resulting $\Delta\alpha=0.285$
from their proposed value $\alpha'=0.786\pm0.398$ one obtaines
$\alpha_{B}=0.501\pm0.398$. \citet{Altavilla:etal:04} have taken
$\Delta\beta=0.045$, $\Delta\gamma=+0.068$, and ${\cal R}_{B}=3.5$;
this leads to $\Delta\alpha=0.198$ which, subtracted from their
preferred value $\alpha'=1.061\pm0.154$, yields
$\alpha_{B}=0.863\pm0.154$. These two determinations of $\alpha_{B}$
embrace the adopted value of $\alpha_{B}=0.612\pm0.073$ in
Table~\ref{tab:coeff3fit} within statistics. 
-- If we had set the coefficient $\beta=0.045\pm0.044$ in
equation~(\ref{eq:color:mean:BV}) equal to 0, as
\citet{Parodi:etal:00} did, the present fiducial sample would give
$\alpha_{B}=0.532\pm0.083$ which is the same as their value within the
errors. 

     The correlation of type-Ia SN luminosity and light curve shape
has a considerably history. First suggested by
\citet{Pskovskii:67,Pskovskii:71,Pskovskii:84} it was taken up by
\citet{Barbon:etal:73}. \citet{Phillips:93} used first the decline
rate $\Delta m_{15}$ to parametrize the correlation, however he had to
depend on inaccurate Tully-Fisher and surface brightness fluctuation
distances, which resulted in an overestimate of the slope $\alpha_{B}$
by a factor of $\sim\!4.5$. The steep slope was the reason why
\citet{Sandage:Tammann:95} questioned the correlation altogether. If
Phillips had been correct in the steepness of the Pskovskii effect,
implying a luminosity variation of $2.5\;$mag in $B$ over the full
range of $\Delta m_{15}$, the use of SNe\,Ia as standard candles would
have been substantially compromised. Only when sufficient relative
velocity distances became available the determination of the slope
$\alpha$ began to converge
\citep{Hamuy:etal:96b,Phillips:etal:99,Parodi:etal:00,Sandage:etal:01,Altavilla:etal:04}.
The remaining differences are now explained by the sensitivity of
$\alpha$ on the adopted relation between color and $\Delta m_{15}$
(equations~\ref{eq:color:mean:BV}$-$\ref{eq:color:mean:VI}). 

     From the above it is clear that a positive/negative excess of the
coefficient $\beta$ in equation~(\ref{eq:color:mean:BV}) propagates
into $E(B\!-\!V)_{\rm host}$ and into the absorption values such that
the magnitudes $M^{00}_{\max}$ (or $m^{00}_{\max}$) are too
faint/bright by ${\cal R}\Delta\beta\Delta m_{15}$. Yet it is to be
noted that the magnitudes $M^{00}_{\max15}$, normalized to $\Delta
m_{15}=1.1$, are not affected by this error, because they are
corrected by an additional term $\Delta\alpha\Delta m_{15} = {\cal
  R}\Delta\beta\Delta m_{15}$ of opposite sign. The error compensation
is only exact if $E(B\!-\!V)_{\rm host}$ is determined from the
intrinsic colors given in equation~(\ref{eq:color:mean:BV}). In the
present case also the intrinsic colors from
equations~(\ref{eq:color:mean:BV35} \& \ref{eq:color:mean:VI}) are used
to determine $E(B\!-\!V)_{\rm host}$, but the three determinations
agree so well that it is sufficient here to show that any effect of an
incorrect value of $\beta$ on $M^{00}_{\max15}$ is compensated by the
adopted value of $\alpha$ in Table~\ref{tab:other:dm15R}.

     \citet{Perlmutter:etal:97} have introduced the ``stretch factor'',
$s$, instead of $\Delta m_{15}$ to characterize the light curve
shape. The two parameters $s$ and $\Delta m_{15}$ show a tight
correlation \citep{Jha:02,Altavilla:etal:04}; they are interchangeable for
all practical purposes.

     An alternative method to allow for the correlation between
luminosity and light curve shape and to determine at the same time the
total absorption $A_{V}$ was introduced by \citet[][see also
\citealt{Jha:02}; \citealt{Wang:etal:03};
\citealt{Tonry:etal:03}]{Riess:etal:96,Riess:etal:98}. Their various
forms of ``Multi-color Light Curve Shape'' (MLCS) fitting are
difficult to apply and offer no obvious advantages.

\subsection{The Dependence of the SN Luminosity on the Color}
\label{sec:other:color}
\begin{figure*}[t]
     \epsscale{0.97}
     \plotone{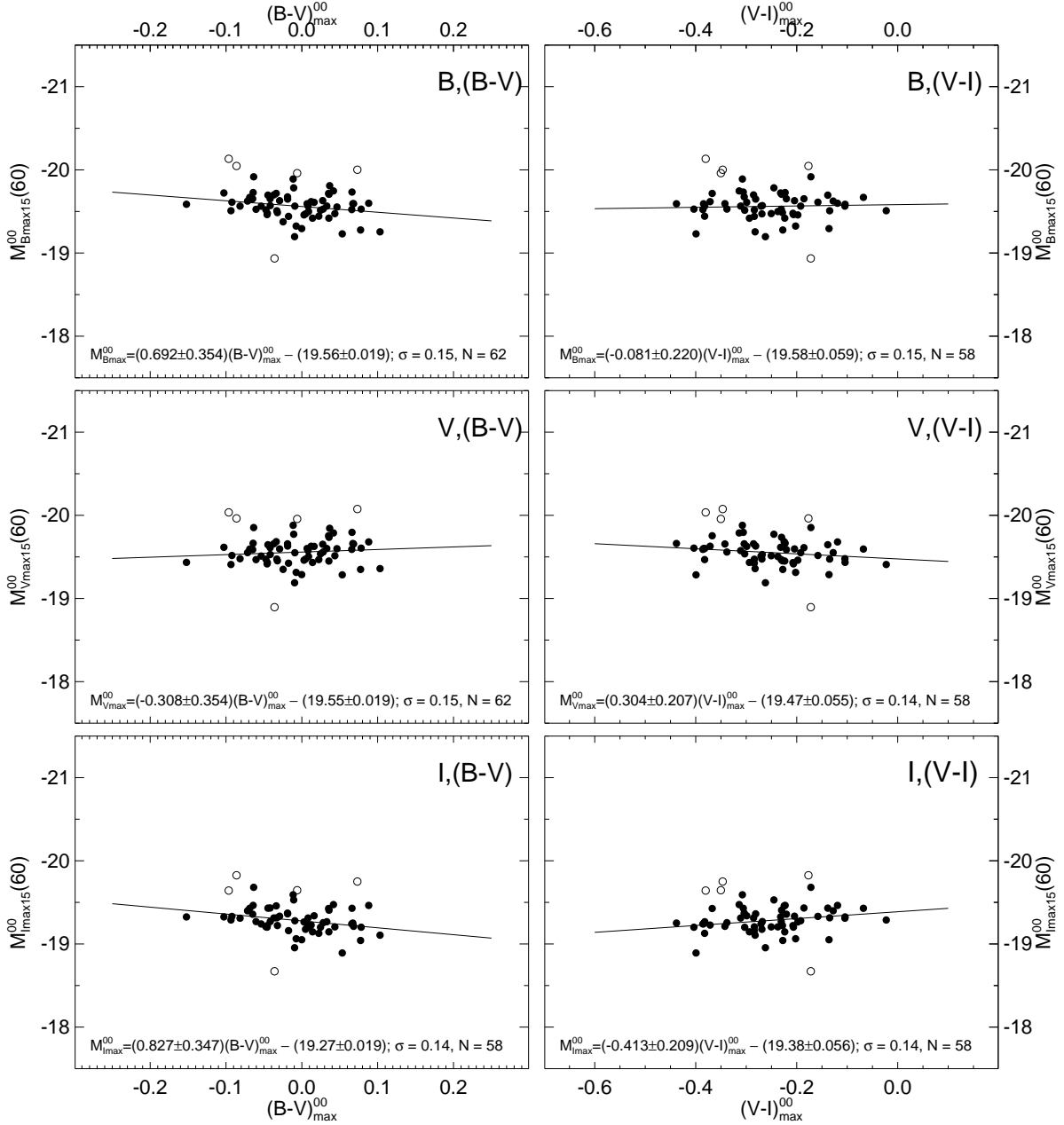} 
     \caption{The dependence of the absolute magnitudes
     $M_{BVI}^{00}$ on the color $(B\!-\!V)_{\max}^{00}$ and
     $(V\!-\!I)_{\max}^{00}$. The SNe\,Ia with open symbols have not
     been used for the solution. They are identified in
     Figure~\ref{fig:absorption:Mabs}.}
\label{fig:other:color}
\end{figure*}

Following a proposal by \citet{Tripp:98} and \citet{Parodi:etal:00} we
test the question whether SN luminosity depends on the color
$(B\!-\!V)^{00}_{\max}$ and/or
$(V\!-\!I)^{00}_{\max}$ in Figure~\ref{fig:other:color}. 
Dependencies at the $\ga2\sigma$ level 
seem to exist of $M^{00}_{B\max15}$ and $M^{00}_{I\max15}$ on
$(B\!-\!V)^{00}_{\max}$ and of $M^{00}_{I\max15}$  on
$(V\!-\!I)^{00}_{\max}$ (see equations at the panel bottoms of
Figure~\ref{fig:other:color}). The correlations should be considered,
however, with some reservations because the range of about
$\pm0.15\;$mag in either color is only about two times the estimated
observational error of a single color determination (about
$\pm0.07\;$mag). 

     In the following the correlation of SN luminosity on
$(B\!-\!V)^{00}_{\max}$ is taken at face value. $(B\!-\!V)$ colors are
preferred over $(V\!-\!I)$ colors because the latter are not available
for all SNe\,Ia of the fiducial sample. The allowance for the color
dependence does not shift the mean luminosity of SNe\,Ia, but brings a
marginal decrease of the luminosity dispersion, at least in $I$. In
any case the allowance for color can bring no harm.  

\subsection{SN Magnitudes Corrected for \boldmath{$\Delta m_{15}$} and Color}
\label{sec:other:corrected}
From the discussion in \S~\ref{sec:other:lightcurve} and
\ref{sec:other:color} follows that minimum scatter of the
absolute magnitude $M^{00}_{BVI}$ can be achieved by
correcting for the decline rate $\Delta m_{15}$ {\em and\/} color
$(B\!-\!V)_{\max}^{00}$. 
The corresponding ansatz
\begin{equation}\label{eq:other:3fitd}
   M_{BVI}^{\rm 00} = a(\Delta m_{15}-1.1) + b[(B\!-\!V) +0.024] + c
\end{equation}
leads with the fiducial sample of 62 SNe\,Ia to the coefficients $a$,
$b$, and $c$ in Table~\ref{tab:coeff3fit}. 
\def\baselinestretch{1.1}
\begin{deluxetable}{lrrrrr}
\tablewidth{0pt}
\tabletypesize{\footnotesize}
\tablenum{5}
\tablecaption{The coefficients in
  equation~(\ref{eq:other:3fitd}).\label{tab:coeff3fit}} 
\tablehead{
& \colhead{$a$} & \colhead{$b$} & \colhead{$c$} & \colhead{$\sigma$} &
\colhead{N}
}
\startdata
\multicolumn{6}{c}{with $(B\!-\!V)_{\max}^{00}$} \\
\noalign{\smallskip}
\tableline
\noalign{\smallskip}
$M_{B}, (B\!-\!V)$: & $0.612\pm0.073$ &  $0.692\pm0.357$ & $-19.57\pm0.02$ & $0.15$ & 62 \\
$M_{V}, (B\!-\!V)$: & $0.612\pm0.073$ & $-0.308\pm0.357$ & $-19.55\pm0.02$ & $0.15$ & 62 \\
$M_{I}, (B\!-\!V)$: & $0.439\pm0.072$ &  $0.827\pm0.350$ & $-19.29\pm0.02$ & $0.14$ & 58 \\
\enddata
  \def\baselinestretch{1.5}
\end{deluxetable}

\begin{figure*}[t]
     \epsscale{1.0}
     \plotone{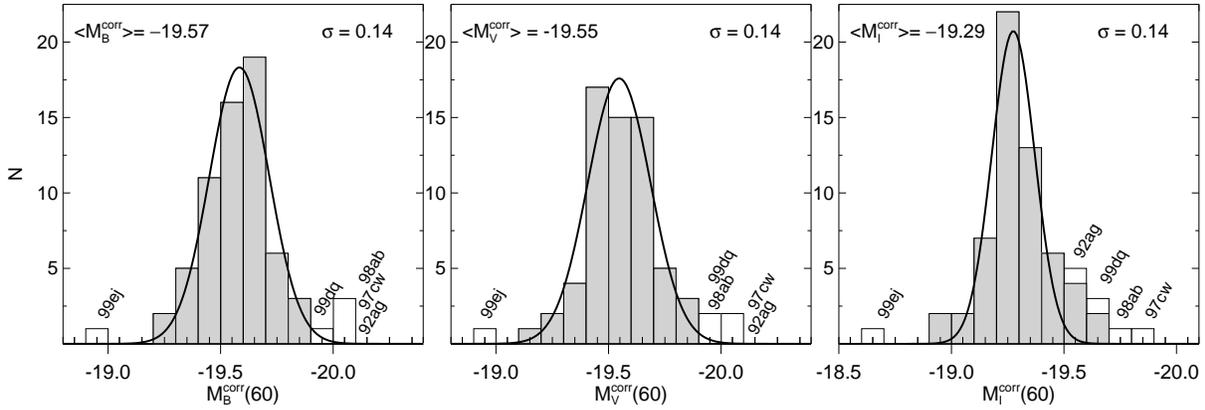}
     \caption{The distribution in absolute magnitude $M_{BVI}^{\rm
     corr}$ of the 62\,(58) SNe\,Ia with
     $3000<v<20\,000\kms$. Additional five SNe\,Ia as in
     Figure~\ref{fig:absorption:Mabs} are 
     identified; they are not used for the solution.}  
\label{fig:MabsHisto}
\end{figure*}

     If one subtracts the first two terms of
equation~(\ref{eq:other:3fitd}) from the fully absorption-corrected
magnitudes $m^{00}_{BVI}$ in columns~5-7 Table~\ref{tab:SN3} one
obtains the magnitudes  $m^{\rm corr}_{BVI}$ (columns 8-10) which are
in addition reduced to $\Delta m_{15}=1.1$ and a corresponding color
of $(B\!-\!V)^{00}_{\max}=-0.024$. The distribution functions of the
resulting absolute magnitudes $M^{\rm corr}_{BVI}$ of the
fiducial sample -- always on the basis of $H_{0}=60$ -- are shown in
Figure~\ref{fig:MabsHisto}. They are well fitted by Gaussians after
five SNe\,Ia, discussed in \S~\ref{sec:absorption:host}, are
excluded. 

\section{PHOTOMETRIC PROPERTIES OF PECULIAR SNe\,Ia (SNe-91T AND
  SNe-91bg AND TWO OTHERS)}
\label{sec:SNpec}
%
\subsection{SNe-91T and SNe-99aa}
\label{sec:SNpec:T}
SNe-91T, with SN\,1991T as the prototype and long suspected to be
overluminous, are spectroscopically well defined \citep[e.g.][and
references therein]{Li:etal:01b}. The 
observed and derived photometric properties of the known quartet are
listed in Table~\ref{tab:SN1}b and \ref{tab:SN3}b, respectively. 
They are characterized by quite slow decline rates $\Delta m_{15}$.
Their intrinsic color $(B\!-\!V)^{00}_{\max}$, being unknown, was 
{\em assumed\/} to be given by equation~(\ref{eq:color:mean:BV}) like
that of normal SNe\,Ia. If one derives color excesses $E(B\!-\!V)_{\rm
  host}$ and absorption values on this basis, one finds the mean
photometric parameters given in Table~\ref{tab:SNeIa-T}. 
The fact that the resulting mean value of
$<\!(V\!-\!I)_{\max}^{\rm corr}\!> = -0.29$ is similar to
normal SNe\,Ia shows that the present assumption on
$(B\!-\!V)_{\max}^{00}$ leads at least to consistent
results. Their absolute magnitudes $M_{BVI}^{00}$, as given
in Table~\ref{tab:SNeIa-T}, turn out to be brighter than normal
SNe\,Ia (equation~\ref{eq:other:Mabs} below) by $\Delta
M_{B}=-0.40\pm0.08$ on average with remarkably small scatter. Even if
one reduces their luminosity to $\Delta m_{15}=1.1$
(equation~\ref{eq:absorption:dm15}) they remain overluminous by
$-0.32\pm0.08$. This conclusion could only be avoided by postulating
that SNe-91T were very blue at maximum, i.e.\
$(B\!-\!V)^{00}_{\max}\approx-0.11$. This is, however, not supported
by interstellar absorption lines which give
$(B\!-\!V)^{00}_{\max}=-0.04$ to $+0.06$ for SN\,1991T
\citep{Filippenko:etal:92} and $(B\!-\!V)^{00}_{\max}=-0.03$ for
SN\,1997br \citep{Li:etal:99}. 
\def\baselinestretch{1.1}
\begin{deluxetable}{lcrrrrrr}
\tablewidth{0pt}
\tabletypesize{\footnotesize}
\tablenum{6}
\tablecaption{Mean photometric parameters of SNe-91T, SNe-99aa, and
  SN\,2000cx (not reduced to a common $\Delta m_{15}$ or
  color).\label{tab:SNeIa-T}} 
\tablehead{
   &  & \colhead{$\Delta m_{15}$} & 
        \colhead{$(B\!-\!V)_{\max}^{00}$} & 
        \colhead{$(V\!-\!I)_{\max}^{00}$} & 
        \colhead{$M_{B\max}^{00}$} &  
        \colhead{$M_{V\max}^{00}$} & 
        \colhead{$M_{I\max}^{00}$} \\ 
  & & \colhead{(1)} & \colhead{(2)} & \colhead{(3)} & \colhead{(4)} &
     \colhead{(5)} & \colhead{(6)}
} 
\startdata
SNe-91T  & mean &   $0.95$ & $-0.03$\tablenotemark{a} &  $-0.29$ & $-20.01$\tablenotemark{b} & $-19.98$\tablenotemark{b} &  $-19.63$\tablenotemark{b} \\
           &      & $\pm.03$ &              & $\pm.06$ & $\pm.08$      & $\pm.08$      &  $\pm.12$      \\
           & N    & 4        & 4            & 3        & 4             & 4             &  3             \\
\noalign{\smallskip}
\multicolumn{8}{c}{-- -- -- -- -- -- -- -- -- -- -- -- -- -- -- -- --
           -- -- -- -- -- -- -- -- -- -- -- -- -- -- -- -- -- -- -- --
           -- -- -- -- -- -- -- -- -- --} \\
\noalign{\smallskip}
SNe-99aa & mean & 0.92     & $-0.01$\tablenotemark{c} & $-0.34$  & $-19.88$      & $-19.87$      &  $-19.53$      \\
           &      & $\pm.03$ &   $\pm0.01$            & $\pm.05$ & $0.10$        & $\pm.10$      &  $\pm.10$      \\
           & N    & 7        & 7            & 7        & 7             & 7             &  7             \\
\noalign{\smallskip}
\multicolumn{8}{c}{-- -- -- -- -- -- -- -- -- -- -- -- -- -- -- -- --
           -- -- -- -- -- -- -- -- -- -- -- -- -- -- -- -- -- -- -- --
           -- -- -- -- -- -- -- -- -- --} \\
\noalign{\smallskip}
SN\,2000cx &      &   $0.97$ & $-0.03$\tablenotemark{a} &  $-0.61$ & $-20.39$      & $-20.36$      &  $-19.75$      \\
           &      &   or     & $ 0.09$\tablenotemark{d} &  $-0.46$ & $-19.96$      & $-20.05$      &  $-19.59$      \\  
\enddata
\tablenotetext{a}{$(B\!-\!V)_{\max}^{00}$ assumed to be
  $-0.030$ from equation~(\ref{eq:color:mean:BV})}
\tablenotetext{b}{For SN\,1991T the Cepheid distance of
  $(m-M)^{0}=30.74$ was adopted \citep{Saha:etal:01}}
\tablenotetext{c}{On the assumption that $(B\!-\!V)^{00}_{\max}$
  and $(B\!-\!V)^{00}_{35}$ are given by
  equation~(\ref{eq:color:mean:BV} \& \ref{eq:color:mean:BV35}) for
  normal SNe\,Ia}
\tablenotetext{d}{Allowing only for Galactic absorption and
  assuming zero reddening in the S0 host galaxy}
  \def\baselinestretch{1.5}
\end{deluxetable}

     SNe-91T are rare events. Four representatives out of a total of
124 corresponds to a relative frequency of about $3\%$. A few
unrecognized SNe-91T may have to be added, but on the other hand
their presumed overluminosity enhances their discovery chance, which
is further enhanced by their slow decline rates. Hence a rate of
roughly $3\%$ seems realistic.

     \citet{Li:etal:01b} have noted in the pre-maximum spectra of some
SNe\,Ia features typical for normal SNe\,Ia and others typical for
SNe-91T \citep[see also][]{Garavini:etal:04}. At maximum and later
they have normal spectra. The prototype is SN\,1999aa; the class is
here designated as SNe-99aa.  Seven SNe\,Ia in Table~\ref{tab:SN1}
were classified as SNe-99aa by \citet{Li:etal:01b}; they are marked
with a dagger. If one determines their host galaxy extinction on the
assumption that they have $(B\!-\!V)$ colors at maximum and at
$t_{B}=35^{\rm d}$ like normal SNe\,Ia, their average color
$(V\!-\!I)_{\max}^{00}$ becomes roughly the same as for normal SNe\,Ia
as seen in Table~\ref{tab:SNeIa-T}. 
Since SNe-99aa are spectroscopically a milder form of SNe-91T one
expects that their luminosity lies between the latter and normal
SNe\,Ia. This expectation is born out in
Table~\ref{tab:SNeIa-T}. SNe-99aa are brighter by only $\Delta
M_{B}=-0.30\pm0.10$ than normal SNe\,Ia, which is further reduced to
$-0.19\pm0.10$ if normalized to $\Delta m_{15}=1.1$.  

     If overluminosity and strikingly slow decline rates ($\Delta
m_{15}<1.02$) are typical for SNe-99aa, it is noted that
Table~\ref{tab:SN1}a contains 13 additional possible candidates. 
At least three of them are normal SNe\,Ia
(SN\,1997bp, \citealt{Anupama:97} and references therein; 
 SN\,1998ef, \citealt{Li:etal:01b}; 
and SN\,2000E, \citealt{Valentini:etal:03}),
leaving 10 candidates at most. If they are added to the 7 known cases,
the total of known or unrecognized SNe-99aa in Table~\ref{tab:SN1}
is $\le14\%$, i.e.\ less than estimated by \citet{Li:etal:01b}. Their
true frequency in a volume-limited sample may be even lower
considering that their suggested overluminosity and their slow decline
rates favor their discovery. 

\subsection{SNe-91bg}
\label{sec:SNpec:bg}
Seven SNe\,Ia in Table~\ref{tab:SN1}b (excluding SN\,1986G) are of
type SNe-91bg with SN\,1991bg as the prototype. They are
characterized by peculiar spectra \citep[][and references
therein]{Mazzali:etal:97}, very red color, low 
luminosity and unparallelledly fast decline rates $\Delta m_{15}$. Their
rate of occurrence seems high in early-type galaxies. This offers a
handle on their intrinsic properties, because the four known
SNe-91bg in early-type galaxies may be assumed to have minimum
reddening. Since the remaining three SNe-91bg in spirals are not
significantly redder [$\Delta(B\!-\!V)^{00}_{\max}=0.66\pm0.09$ versus
$0.59\pm0.08$] we assume that the internal reddening of this class can
be neglected. Minimal reddening is also independently confirmed for
SN\,1999bg in the spiral NGC\,2841 from late-phase multi-color
photometry \citep{Garnavich:etal:04}.
The ensuing mean photometric parameters are compiled in
Table~\ref{tab:SNeIa-bg}. 

\def\baselinestretch{1.1}
\begin{deluxetable}{ccrrrrrrr}
\tablewidth{395pt}
\tablenum{7}
\tabletypesize{\footnotesize}
\tablecaption{Mean photometric parameters of SNe-91bg.\label{tab:SNeIa-bg}} 
\tablehead{
    & & \colhead{$\Delta m_{15}$} & 
        \colhead{$(B\!-\!V)_{\max}^{00}$} & 
        \colhead{$(V\!-\!I)_{\max}^{00}$} & \colhead{~~} &
        \colhead{$M_{B\max}^{00}$} &  
        \colhead{$M_{V\max}^{00}$} & 
        \colhead{$M_{I\max}^{00}$} \\ 
  & & \colhead{(1)} & \colhead{(2)} & \colhead{(3)} & & \colhead{(4)} &
     \colhead{(5)} & \colhead{(6)}
} 
\startdata
mean &&   $1.91$ &   $0.62$     &   $0.23$ && $-17.46$\tablenotemark{a} & $-18.07$\tablenotemark{a} &  $-18.30$\tablenotemark{a} \\
         && $\pm.02$ & $\pm.06$     & $\pm.03$ && $\pm.16$   & $\pm.14$   &  $\pm.11$      \\
$\sigma$ &&   $0.04$ &   $0.15$     &   $0.09$ &&   $0.42$   &   $0.37$   &    $0.30$      \\
N        && 7        & 7            & 7        && 7          & 7          &  7             \\
\noalign{\smallskip}
\multicolumn{9}{c}{-- -- -- -- -- -- -- -- -- -- -- -- -- -- -- -- --
           -- -- -- -- -- -- -- -- -- -- -- -- -- -- -- -- -- -- -- --
           -- -- -- -- -- -- -- -- --} \\
\noalign{\smallskip}
         &&   $1.69$ & $0.00$\tablenotemark{b} & \nodatr  && $-19.16$\tablenotemark{d} & $-19.16$\tablenotemark{d} &  \nodatr \\
   \raisebox{1.5ex}[0cm][0cm]{86G} &&   or     &  $0.43$\tablenotemark{c} & \nodatr  && $-17.58$\tablenotemark{d} & $-18.01$\tablenotemark{d} &  \nodatr \\
\enddata
\tablenotetext{a}{Adopting for SN\,1999by the Cepheid
  distance of $(m-M)^{0}=30.74$ \citep{Macri:etal:01}}
\tablenotetext{b}{$(B\!-\!V)_{\max}^{00}$ assumed to be
  $0.00$ as for all normal SNe\,Ia with $\Delta m_{15}=1.69$ (see
  equation~\ref{eq:color:mean:BV})} 
\tablenotetext{c}{$(B\!-\!V)_{35}^{00}$ assumed to be
  $1.067$ as for normal SNe\,Ia with $\Delta m_{15}=1.69$ (see
  equation~\ref{eq:color:mean:BV35})} 
\tablenotetext{d}{Adopting for the parent galaxy NGC\,5128
  (Cen\,A) $(m-M)^{0}=27.80$ \citep{Thim:etal:03}}
  \def\baselinestretch{1.5}
\end{deluxetable}

     The SNe-91bg emerge -- besides for their spectral peculiarities
-- as a separate class of SNe with quite narrowly confined photometric
properties. Their decline rate varies little about $\Delta
m_{15}=1.91$, while no normal SN\,Ia is known with $\Delta
m_{15}\ga1.7$. Their extremely red intrinsic colors have only moderate
scatter. In fact their $(V\!-\!I)_{\max}^{00}$ shows less scatter than
normal SNe\,Ia. Their absolute magnitudes show somewhat more
variation than normal SNe\,Ia, but they are all strikingly
underluminous by $\Delta M_{B}=2.11\pm0.16$, $\Delta
M_{V}=1.48\pm0.14$, and  $\Delta M_{I}=0.93\pm0.11$ on average. 
The underluminosity may be somewhat reduced if we have underestimated
the host galaxy absorption, but we deem values as large as $A_{B\,{\rm
    host}}\approx 0.3$ (and correspondingly less in $V$ and $I$) as
unlikely. 
Models of two SN-91bg (SN\,1991bg and SN\,1999by) were
elaborated by \citet{Mazzali:etal:97} and by \citet{Hoeflich:etal:02}.

     The frequency per unit volume of SNe-91bg can roughly be
estimated by noting that the seven known cases all lie within
$5000\kms$, and that Table~\ref{tab:SN1}a contains 47 SNe\,Ia within
this distance limit. 
Of these, 45 have $\Delta m_{15}<1.7$ and the remaining two
(SN\,1981D and 1984A) are normal SNe\,Ia on the basis of their
colors. Hence seven cases out of a total of 54 correspond to a
frequency of $13\%$, or somewhat more because their fast decline rates
impair their discovery probability \citep[see also][]{Li:etal:01a}. 
Since they occur preferentially in early-type galaxies, it is
suggested that they stem from an older population \citep{Howell:01}. 

\subsection{Two Additional Peculiar SNe\,Ia (SN\,2000cx AND SN\,1986G)}
\label{sec:SNpec:add}
Only two SNe\,Ia in Table~\ref{tab:SN1} are neither normal nor belong
to the SNe-91T (including SNe-99aa) and SNe-91bg classes.
\subsubsection{SN\,2000cx}
\label{sec:SNpec:add:00ca}
The spectrum of SN\,2000cx resembles SNe-91T at early phases, but
its light curve is highly unusual, in fact unique
\citep{Li:etal:01c,Candia:etal:03}. The observed photometric
parameters are given in Table~\ref{tab:SN1}b. The intrinsic
photometric properties at maximum are calculated here for two
different assumptions. 
a) The SN has $(B\!-\!V)_{\max}^{00}=-0.01$ like normal SNe\,Ia, or 
b) The SN suffers zero reddening in its S0 host galaxy. The ensuing
colors and luminosities are shown in Table~\ref{tab:SNeIa-T}. In
either case SN\,2000cx may be with $M_{B\,\max}^{00}\la-20.0$ one of
the brightest SN in the entire sample.  
\citet{Fisher:etal:99} have argued that such high luminosities,
although still depending on the exact value of $H_{0}$, require
one Chandrasekhar mass of pure $^{56}$Ni which cannot be provided by a
single White Dwarf. They proposed therefore a model with two merging
White Dwarfs. SN\,2000cx may be a prime candidate for this process.

\subsubsection{SN\,1986G}
\label{sec:SNpec:add:86G}
The spectrum of SN\,1986G is similar to SNe-91bg, but less extreme
\citep[][and references therein]{Cristiani:etal:92}. Its photometric
observables are in Table~\ref{tab:SN1}c. They show that SN\,1986G was
observed to be even redder than SNe-91bg, but to have a clearly
slower decline rate. Much of its redness is certainly due to its
position in the dust lane of NGC\,5128 (Cen\,A).

     The intrinsic color and luminosity are calculated here for
two different assumptions.
a) $(B\!-\!V)^{00}_{\max}$ of SN\,1986G is given by
equation~(\ref{eq:color:mean:BV}) like normal SNe\,Ia,
and b) it has the tail color $(B\!-\!V)_{35}^{00}=1.067$ which, from
equation~(\ref{eq:color:mean:BV35}), is the tail color of a normal SN
with $\Delta m_{15}=1.69$. The results for both cases are shown in
Table~\ref{tab:SNeIa-bg}. In the latter case, with the tail color as
standard, the extinction $E(B\!-\!V)_{\max\,35}$ becomes $0.489$,
which seems on the low side given its position in a prominent dust
lane.
However, even the small reddening makes SN\,1986G bluer than average
SNe-91bg and its relatively slow decline rate ($\Delta m_{15}=1.69$)
remains untypical for this  class, although with the distance modulus
of NGC\,5128 of $(m-M)^{0}=27.80$ \citep{Thim:etal:03} its absolute
magnitudes $M_{BV}$ become quite similar to SNe-91bg. On the other
hand case a) leads to $E(B\!-\!V)_{\rm host}=0.922$ and
$E(B\!-\!V)_{\rm total}=1.037$ (including Galactic reddening) in good
agreement with the independent 
estimates of \citet{Rich:87} and \citet{Cristiani:etal:92}. Hence, we
prefer solution a) and conclude that the luminosity of SN\,1986G
lies inbetween normal SNe\,Ia and SNe-91bg (see
Table~\ref{tab:SNeIa-bg}). 

\begin{figure}[t]
     \epsscale{0.49}
     \plotone{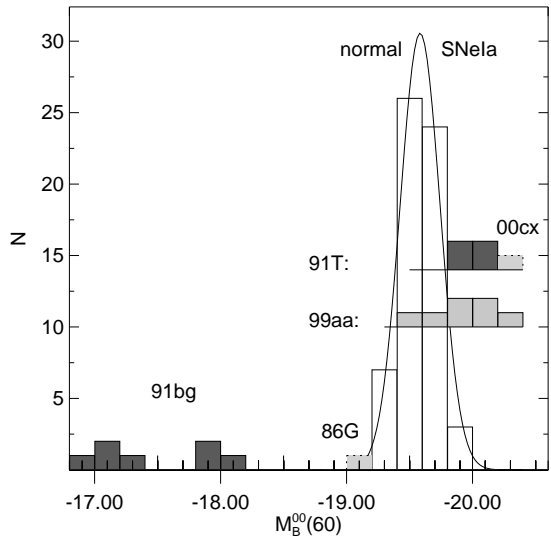}
     \caption{The distribution in absolute magnitude $M^{00}_{B}$ of
     SNe\,Ia. The 58 normal SNe\,Ia of the fiducial sample, normalized
     for $\Delta m_{15}$ and $(B\!-\!V)^{00}_{\max}$, are shown in the
     open histogram with a fitted Gauss curve. The spectroscopically
     peculiar SNe-91bg and SNe-91T are in dark grey, the SNe-99aa in
     light grey. In addition the unusual SN\,1986G and 2000cx are
     shown.}   
\label{fig:SNe-Pec}
\end{figure}

     An overview of the luminosity distribution in $B$ of the various
types of SNe\,Ia is given in Figure~\ref{fig:SNe-Pec}.

\section{THE HUBBLE DIAGRAM OF NORMAL SNe\,Ia}
\label{sec:hubble}
\begin{figure}[t]
     \epsscale{0.53}
     \plotone{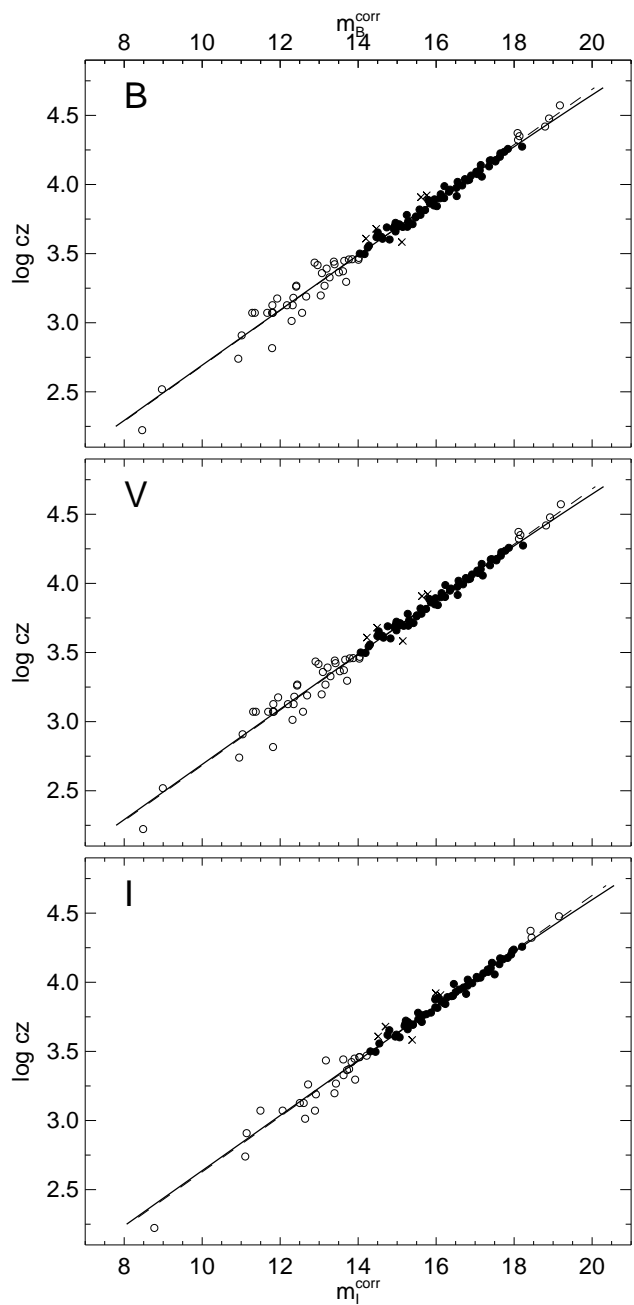}
     \caption{The Hubble diagram in $B$, $V$, and $I$ of 108 normal
     SNe\,Ia. The $\Omega_{\rm M}=0.3, \Omega_{\Lambda}=0.7$ fits 
     to the Hubble line in each color as given in 
     Table~\ref{tab:hubble} do not use five SNe\,Ia (plotted with
     $\times$) discussed in \S~\ref{sec:absorption:host}. The SNe\,Ia
     with $v<3000\kms$ and $v>20\,000\kms$ are shown as open
     symbols. The dashed line corresponds to a flat matter-dominated
     Universe.}  
\label{fig:hubble}
\end{figure}

The Hubble diagram -- $\log cz$ vs. $m_{BVI}^{\rm corr}$ -- is shown in
Figure~\ref{fig:hubble} for 108 normal SNe\,Ia. The recession velocities
are corrected for streaming motions as described in the explanation of
column~5 of Table~\ref{tab:SN1} (\S~\ref{sec:Data}). 
The apparent magnitudes are corrected for
Galactic and host galaxy absorption as laid out in
\S~\ref{sec:absorption} and reduced to $\Delta m_{15}=1.1$
and $(B\!-\!V)_{\max}^{00}=-0.024$ (equation~\ref{eq:other:3fitd}). 

     A slightly curved Hubble line, corresponding to a flat Universe
with $\Omega_{\Lambda}=0.7$, is fitted to the points in
Figure~\ref{fig:hubble}. The corresponding relation is given by
\citet{Carroll:etal:92} 
\begin{equation}\label{eq:hubble:01}
   \phi(z) = 0.2 m_{\lambda}^{\rm corr} + C_{\lambda},
\end{equation}
where
\begin{equation}\label{eq:hubble:02}
   \phi(z) = \log c(1+z_{1})\int_{0}^{z_{1}}[(1+z)^{2}(1+0.3z)-0.7z(2+z)]^{-1/2}dz,
\end{equation}
and where the intercept is given by
\begin{equation}\label{eq:hubble:03}
   C_{\lambda} = \log H_{0} - 0.2 M_{\lambda}^{\rm corr} -5.
\end{equation}
From this follows the decisive relation
\begin{equation}\label{eq:hubble:04}
    \log H_{0} = 0.2 M_{\lambda}^{\rm corr} + C_{\lambda} + 5.
\end{equation}
As equation~(\ref{eq:hubble:04}) shows, the determination of $H_{0}$
requires an independent calibration of $M_{\lambda}^{\rm corr}$ of
normal SNe\,Ia and an evaluation of the intercept $C_{\lambda}$. While
a revised calibration of $M_{\lambda}^{\rm corr}$ will be provided in
the forthcoming summary paper, the determination of $C_{\lambda}$ is
given here.

     Nine solutions for $C_{\lambda}$ are presented in
Table~\ref{tab:hubble} for different subsets of the SNe\,Ia in
Table~\ref{tab:SN1}a. The subsets are defined by velocity cuts and by
different amounts of the total reddening $E(B\!-\!V)_{\rm t}$.

     It is very satisfactory that the solutions 1-3 and 6-8 for
SNe\,Ia beyond $3000\kms$ give the same values for $C_{\lambda}$ in
each waveband within the statistical errors. $C_{B}$ varies by
$\pm0.003$ implying a variation in mean absolute magnitude
$<\!\!M_{B}^{\rm corr}\!\!>$ of only $\pm0.015\mag$, which influences
$H_{0}$ by less than $\pm1\%$.

     Solution~3 is from the fiducial sample as defined in
\S~\ref{sec:absorption:host}. It is least affected  by photometric
errors, peculiar motions, possibly remaining 
errors of the K-correction, and cosmological effects. The
corresponding values of $C_{BVI}$ with equation~(\ref{eq:hubble:03})
and $H_{0}=60$ give 
\begin{equation}\label{eq:other:Mabs}
   M_{B}^{\rm corr} = -19.57\pm0.02,\quad
   M_{V}^{\rm corr} = -19.55\pm0.02, \mbox{ and}\quad
   M_{I}^{\rm corr} = -19.29\pm0.02.
\end{equation}
It is also
noted that $C_{B}-C_{V}=0.005\pm0.006$ corresponding to 
$(B\!-\!V)^{00}=-0.025\pm0.03$, and that
$C_{V}-C_{I}=0.051\pm0.006$ corresponding to
$(V\!-\!I)^{00}=-0.26\pm0.03$, -- both values being consistent with
the observed colors in equation~(\ref{eq:color:mean}).

\def\baselinestretch{1.1}
\begin{deluxetable}{lcccccccc}
\tablewidth{0pt}
\tablenum{8}
\tabletypesize{\scriptsize}
\tablecaption{Nine solutions of $C_{\lambda}$ in equation~(\ref{eq:hubble:04})
  for different choices of the SNe\,Ia sample.\label{tab:hubble}} 
\tablehead{
    & \colhead{$C_{B}$} & \colhead{$\sigma_{B}$}
    & \colhead{$C_{V}$} & \colhead{$\sigma_{V}$} & \colhead{$\!\!N_{BV}\!\!\!\!$}
    & \colhead{$C_{I}$} & \colhead{$\sigma_{I}$} & \colhead{$N_{I}$}
}
\startdata
1) all:               & $0.696\pm0.006$ & $0.30$ & $0.691\pm0.006$ & $0.30$ & $105$ & $0.634\pm0.006$ & $0.27$ & $85$  \\ 
2) $v>3000[\kms]$:    & $0.697\pm0.004$ & $0.16$ & $0.693\pm0.004$ & $0.16$ &  $68$ & $0.640\pm0.004$ & $0.15$ & $61$  \\ 
3) $3000<v<20000$:    & $0.693\pm0.004$ & $0.14$ & $0.688\pm0.004$ & $0.14$ &  $62$ & $0.637\pm0.004$ & $0.14$ & $58$  \\ 
4) $v<2000$:          & $0.675\pm0.021$ & $0.51$ & $0.670\pm0.021$ & $0.51$ &  $23$ & $0.590\pm0.025$ & $0.47$ & $14$  \\ 
5) $v<1200$:          & $0.656\pm0.034$ & $0.59$ & $0.651\pm0.034$ & $0.59$ &  $12$ & $0.583\pm0.045$ & $0.60$ & $ 7$  \\ 
6) $v>3000,E_t<0.30$: & $0.697\pm0.004$ & $0.16$ & $0.692\pm0.004$ & $0.16$ & $ 63$ & $0.639\pm0.004$ & $0.15$ & $58$  \\ 
7) $v>3000,E_t<0.15$: & $0.697\pm0.005$ & $0.17$ & $0.693\pm0.005$ & $0.17$ &  $51$ & $0.638\pm0.005$ & $0.15$ & $46$  \\ 
8) $v>3000,E_t<0.05$: & $0.692\pm0.006$ & $0.16$ & $0.687\pm0.006$ & $0.16$ &  $24$ & $0.635\pm0.006$ & $0.15$ & $21$  \\ 
9) $v>5000$, Table\,2:& $0.695\pm0.007$ & $0.17$ & $0.692\pm0.006$ & $0.14$ &  $21$ & $0.640\pm0.005$ & $0.10$ & $20$  \\
\enddata
  \def\baselinestretch{1.5}
\end{deluxetable}

     The solutions 6-8 admit different amounts of the total
extinction $E(B\!-\!V)_{\rm t}$ in the Galaxy and in the host
galaxy. This has virtually no effect on $C_{I}$, and even $C_{B}$,
which is most sensitive to absorption corrections,
changes by not more than $0.005$ (corresponding to $0.025\mag$). This
proves in 
favor of the adopted unconventional values of ${\cal R}_{BVI}$
(equation~\ref{eq:absorption:R}). The SNe\,Ia in Table~\ref{tab:SN1}a have
$<\!\!E(B\!-\!V)_{\rm t}\!\!>=0.16$, 
and an error of ${\cal R}_{B}$ of more than $\Delta {\cal
  R}_{B}\approx\pm0.25$ would cause a larger discrepancy.
The limitation of the maximum value of $E(B\!-\!V)_{\rm t}$ does not
lead to a decrease of the scatter about the Hubble line. 
This suggest that the magnitude dispersion is mainly due to {\em
  random\/} errors of $E(B\!-\!V)_{\rm t}$ and of the corresponding
absorption corrections.

     If one averages the $B$, $V$, and $I$ magnitudes to define the
Hubble line in solutions 1-8, the scatter is {\em not\/} reduced,
which indicates that the magnitude errors are strongly correlated.

     It is not meaningful to compare the values of $C_{BVI}$
derived here with those of other authors, because they depend on
a number of choices taken here: the intrinsic color of SNe\,Ia, the
reddening and absorption law, and the reduction to the standard values
of $\Delta m_{15}=1.1$ and $(B\!-\!V)_{\max}^{00}=-0.024$. 
It is therefore mandatory, if the present values of $C_{\lambda}$ are
to be used to obtain $H_{0}$ through equation~(\ref{eq:hubble:04}),
that the calibrating SNe\,Ia are treated according to the
prescriptions given here. 

     The dispersion about the Hubble line in solution~3, i.e.\
$\sigma_{BVI}=0.14\mag$, is quite satisfactory
for a large sample of 62 (58) SNe\,Ia. 
\citet{Tripp:98} has found from 29 SNe\,Ia $\sigma_{B}=0.15\mag$, but
his reddening-to-absorption ratio of ${\cal R}_{B}=2.09$ for Galactic
and host galaxy absorption seems unrealistic.
Previous values of 
$\sigma_{V}=0.14$ \citep{Hamuy:etal:96b,Phillips:etal:99} or even 
$\sigma_{V}=0.12\mag$ \citep{Parodi:etal:00} are based on
smaller samples, and are in addition color-limited by
$(B\!-\!V)_{\max}\le0.2$. \citet{Altavilla:etal:04} have found
$\sigma_{B}=0.20\mag$ for 18 of their low-extinction SNe\,Ia. For
comparison the MLCS method of fitting the SN light curves has yielded 
$\sigma_{V}=0.16$ \citep{Riess:etal:99,Jha:etal:99} and
$\sigma_{V}=0.20\mag$ \citep{Tonry:etal:03}. The ``Bayesian Adapted
Template Match'' (BATM) of \citet{Barris:Tonry:04} yields a scatter of
$\sigma_{(m-M)}=0.21\mag$. The very small scatter of $\sigma=0.08\mag$
achieved by \citet{Wang:etal:03} with their ``Color-Magnitude Intercept
Calibration'' may not be representative because it is derived from a
severely color-restricted sample; their color limit
$(B\!-\!V)_{\max}^{0}\le0.05$ cuts into the realistic color
distribution of even unreddened SNe\,Ia (cf.\
Figure~\ref{fig:color:tab2}).

     A physically more relevant question concerns the {\em
intrinsic\/} scatter of homogenized SN luminosities. The best handle
for this is offered by the 21 (20) SNe\,Ia in
Table~\ref{tab:color:minred} with minimum absorption in their host
galaxies and $v>5000\kms$ (solution~9 of Table~\ref{tab:hubble}). The
higher velocity limit is chosen here to guard against peculiar motions
which, if $\sim\!300\kms$, can contribute a magnitude dispersion of
$0.2\;$mag still at $v=3000\kms$. The sample gives a puzzlingly large
value of $\sigma_{B}=0.17$, but quite low values for $\sigma_{V}=0.14$
and $\sigma_{I}=0.10\mag$. 
The small scatter in $I$ is particularly noteworthy because $I_{\max}$,
occurring before $B_{\max}$, requires good photometry at an early
phase. Allowing conservatively for an average photometric error of
$\sigma_{m}=0.05\mag$, we derive an intrinsic scatter of the
homogenized $I_{\max}$ magnitudes of $\la0.10\mag$, which -- in view
of the many small and not so small spectroscopic and light curve
differences between normal SNe\,Ia -- is remarkably small. 

     We have refrained from weighting the SN data, as some
authors have done, by the quoted magnitude errors or by the time
interval between inferred $B$ maximum and the onset of the photometry,
although these data are given for most, but not all SNe\,Ia in
Table~\ref{tab:SN1}a. The reason is mainly that the quoted magnitude
errors from many different sources carry an subjective element, and
that they include sometimes systematic errors and sometimes not. We
have instead made alternative solutions excluding SNe\,Ia with large
quoted magnitude errors and/or with a late onset of their
photometry. In no case the result or even the statistical error was
significantly changed. In view of the large sample considered here
this is no surprise.

     A last comment to the solutions in Table~\ref{tab:hubble} concerns
the relative size of $H_{0}$ in the local Universe. The strong
increase of $\sigma_{BVI}$ in solution~4 and 5 is certainly due to
peculiar motions whose influence becomes important at small redshifts.
The median velocity of the SNe\,Ia in solution~4 is $1200\kms$.
The observed magnitude scatter of $\sigma_{I}=0.47$ corresponds
therefore to typical peculiar motions of $\sim\!280\kms$. Within
$v<1200\kms$ the median velocity is $1000\kms$ and the scatter
corresponds here to $\sim\!300\kms$. 
These values are high relative to the small rms peculiar velocity 
of $<100\kms$ consistently derived over still smaller scales 
\citep[e.g.][]{Tammann:Sandage:85,Sandage:86,Sandage:87,Karachentsev:Makarov:96}, 
but many of the SNe\,Ia lie in the Virgo cluster region which is known
to be noisy \citep{Tammann:etal:02}, and peculiar velocities are
generally expected to increase with increasing scale length up to some
asymptotic value which, judging from the CMB dipole, is
$\ga630:\sqrt{3}=360\kms$ in the radial direction. -- 
For $v<2000\kms$ $C_{BV}$ becomes smaller by $0.018\pm0.021$ and
$C_{I}$ by $0.047\pm0.025$ than in solution~3. This correspond from
equation~(\ref{eq:hubble:04}) to a local decrease of $H_{0}$ by $5\pm5\%$
or $12\pm6\%$. Analogously the very local value of $H_{0}$ is, from
solution~5, $9\pm9\%$ or $16\pm14\%$ smaller than the cosmic
value. The local decrease of $H_{0}$ is in view of the small number of
SNe\,Ia and the statistical errors only suggestive. 

     The mean absolute magnitude $M^{\rm corr}_{V}$ in
equation~(\ref{eq:other:Mabs}) is combined with the individual
apparent magnitudes $m^{\rm corr}_{V}$ in Table~\ref{tab:SN3}a,
column~9, to yield photometric absorption-free distance moduli for 111
normal SNe\,Ia and hence of their parent galaxies (Table~\ref{tab:SN3}a,
column~11). The expected random error is only
$\sigma_{m\!-\!M}=0.14\mag$; some moduli suspected for one reason or
another to be less accurate are shown in parentheses. The moduli have
the advantage -- particularly for the nearby galaxies -- to be
independent of peculiar motions. Of course, all distances are still to
be shifted by a constant amount, once the adopted value of $H_{0}=60$
can be replaced by the actual value of $H_{0}$, which is the ultimate
aim of this series.

\section{SUMMARY AND CONCLUSIONS}
\label{sec:conclusions}
The main purpose of this paper is to update an earlier 
summary by \citet{Parodi:etal:00} on the reddening, absorption and 
second parameter corrections for a color-restricted sample of "Branch 
normal" type Ia supernovae where the corrected data were used 
to construct the Hubble diagram in $B$, $V$, and $I$. Here the sample 
is increased to 124 type Ia supernovae with no restrictions 
placed on the observed color, thereby avoiding the criticisms of 
the \citeauthor{Parodi:etal:00} sample raised by others concerning
observational selection biases.  

     All SN believed to be "Branch normal" \citep{Branch:etal:93}  
were used, together with 13 spectroscopically peculiar SNe\,Ia 
(types 91T, 91bg, SN\,2000cx, 1986G) that deviate in one way or another 
from the main sample. The peculiar supernovae are listed 
separately in Tables~\ref{tab:SN1}b, \ref{tab:SN1}c, \ref{tab:SN3}b,
and \ref{tab:SN3}c and are discussed in \S~\ref{sec:SNpec}.
     
     Because no a priori cut in observed color at maximum was  
imposed, many of the SNe in our enlarged sample 
suffer non-negligible reddening in their host galaxies, and care   
has been taken at each step in the reductions to guard against 
systematic errors in the derived absorptions. 

     The twelve main research points made in the paper are these. 

     (1) The 22 SNe\,Ia in E or S0 galaxies and the 12 SNe\,Ia 
in the outlying regions of spirals, set out in
Table~\ref{tab:color:minred}, are expected to have nearly negligible
extinction due to the host galaxy. 
Their mean intrinsic colors, corrected for Galactic reddening and
nearly free of any host galaxy reddening, are
$(B\!-\!V)^{00}_{\max}=-0.013$, $(B\!-\!V)^{00}_{35}=1.092$ (i.e.\ the
tail color at phase $t_{B}=35^{\rm d}$),  and
$(V\!-\!I)^{00}_{\max}=-0.232$. The individual colors
$(B\!-\!V)^{00}_{\max}$ become marginally redder, the colors
$(V\!-\!I)^{00}_{\max}$ significantly redder with increasing   
decline rate $\Delta m_{15}$, while the tail colors
$(B\!-\!V)^{00}_{35}$ become clearly bluer with increasing $\Delta
m_{15}$ (equations~\ref{eq:color:mean:BV}$-$\ref{eq:color:mean:VI}). The
latter result is in variance with  
\citet{Lira:95}, 
\citet{Phillips:etal:99},
\citet{Jha:02}, and 
\citet{Altavilla:etal:04},
who assumed the tail colors to be independent of any second
parameters. 

     (2) The host galaxy reddenings for all other normal SNe\,Ia in
the sample are based on the intrinsic colors derived in item~(1)
above. The difference between the observed, Galactic-reddening-free
colors $(B\!-\!V)^{0}_{\max}$, $(B\!-\!V)^{0}_{35}$, and
$(V\!-\!I)^{0}_{\max}$ and the corresponding colors from 
equations~(\ref{eq:color:mean:BV}$-$\ref{eq:color:mean:VI}), i.e.\
adjusted to the appropriate value of $\Delta m_{15}$, yield three
different estimates of the reddening $E(B\!-\!V)_{\max}$,
$E(B\!-\!V)_{35}$, and $E(V\!-\!I)_{\max}$.    

     (3) Due to the general color dependence of color excesses caused
by the photometric bandwidth effect, the excesses $E(B\!-\!V)_{35}$,
determined in item~(2) from a late, quite red 
phase of the SN light curves, must be converted to the value
applicable at the blue maximum phase. This is achieved by requiring
that $(B\!-\!V)^{00}_{\max}$ as well as $(V\!-\!I)^{00}_{\max}$ be
independent of $E(B\!-\!V)_{35}$
(cf. Figures~\ref{fig:color:E35} \& \ref{fig:color:EVIhost}). 
The requirement leads to equation~(\ref{eq:color:EBVmax35}) which
converts $E(B\!-\!V)_{35}$ to $E(B\!-\!V)_{max35}$, and to
equation~(\ref{eq:color:EBVVIhost}) which converts $E(V\!-\!I)$ into
$E(B\!-\!V)^{V\!-\!I}_{\max}$. The three determinations of
$E(B\!-\!V)_{\max}$ are intercompared in Figures~\ref{fig:color:comp1}
\& \ref{fig:color:comp2}. They agree well and are averaged to give the
adopted reddening $E(B\!-\!V)_{\rm host}$ in the host galaxy. The
random error of $E(B\!-\!V)_{\rm host}$ is $0.043\;$mag.

     (4) The adopted color excesses $E(B\!-\!V)_{\rm host}$ depend on
the assumption  that the sample of SNe\,Ia in E/S0 galaxies and the
outliners in spirals in Table~\ref{tab:color:minred} suffer no
host-galaxy reddenings. 
If, in fact, they do suffer a non-zero amount of
reddening, the true intrinsic colors of the complete sample,
determined by the method of items~(1) through (3), will be bluer, and
the magnitudes brighter than adopted here, but this will have no
effect on the calibration of $H_{0}$ to be made in the final summary
paper of this series as long as the calibrating SNe\,Ia in host
galaxies are reduced to the {\em same\/} color. 
In that case, the local calibrators would suffer the same error in 
absorption as would the distant SNe\,Ia defining the global
Hubble diagram, and the error cancels.
   
     (5) The magnitudes $m^{0}_{BVI\max}$, corrected for conventional
Galactic absorption, were transformed into absolute magnitudes
$M^{0}_{BVI\max}$ for a ``fiducial'' sample of 62\,(58) SNe\,Ia
which have $3000<v_{\rm CMB}<20\,000\kms$ (excluding only the highly
absorbed SN\,1996ai and five SNe\,Ia deviating by more than $2\sigma$
and discussed in \S~\ref{sec:absorption:host}) to guard against
peculiar motions and other effects. A flat Universe with $H_{0}=60$,
$\Lambda=0.7$ was assumed. We show in \S~\ref{sec:absorption:host} a
strong correlation of $M^{0}_{BVI\max}$ with $E(B\!-\!V)_{\rm host}$
(Figure~\ref{fig:absorption:Mabs}).  
This correlation determines the ratio of absorption-to-reddening in
the host galaxy. The absorption-to-reddening values determined in this
way are 
${\cal R}_{B}=3.65\pm0.16$, 
${\cal R}_{V}=2.65\pm0.15$, and
${\cal R}_{I}=1.35\pm0.21$.
These well determined values differ significantly from 
the canonical values in the Galaxy of ${\cal R}_{BVI}$ of
$4.1$, $3.1$, and $1.8$. The significant difference with the
Galactic values has been pointed out before by 
\citet{Branch:Tammann:92},
\citet{Riess:etal:96},
\citet{Phillips:etal:99}, 
\citet{Krisciunas:etal:00},
\citet{Wang:etal:03}, and 
\citet{Altavilla:etal:04}. 
An explanation is that the intense radiation of the supernovae in some
manner modifies the size distribution of the dust grains in the host
galaxy near the SN.
The canonical values of ${\cal R}_{BVI}$ have been used to determine
the Galactic absorption and the new, smaller values of ${\cal
  R}_{BVI}$ for the absorption in the host galaxy.
 
     (6) The absorption-free magnitudes $M^{00}_{V}$ of the fiducial
sample are plotted against the Hubble type of the host galaxies in
Figure~\ref{fig:other:type}. It is confirmed that the 
supernovae in  E/S0 galaxies average $\sim\!0.3$ magnitude
fainter than SNe in late-type spirals,
known already since 1995  
(\citealt{Hamuy:etal:95};  
\citealt{Saha:etal:97,Saha:etal:99};
\citealt{Sandage:Tammann:97}; 
\citealt{Sandage:etal:01}, Figure~6). 
It has been seen in a different form by the discovery by 
\citet[][their Figure~4]{Hamuy:etal:96b}, that the decline rate, 
$\Delta m_{15}$, is correlated  
with galaxy type (SNe\,Ia in E/S0 galaxies have average decline rates
near $1.5\mag$, while those in late-type spirals have a mean rate near
$1.0$), coupled with the fact that the SNe\,Ia with large decline rates 
are fainter than those with smaller rates. 

     The reason for the $M^{00}_{\max}$-galaxy type correlation 
is expected to be related to the different gestation 
times for type Ia SNe and the subsequent difference in the Fe 
abundance between E/S0 galaxies and spirals of the same age. The 
consequences may or may not be profound in using SNe\,Ia for 
cosmological probes as is the current trend 
\citep{Riess:etal:98,Riess:etal:04,Perlmutter:etal:99,Knop:etal:03,Barris:etal:04},
depending on the precision 
with which the effect can be compensated for by the $\Delta m_{15}$
correlations with galaxy type.
  
     (7) The fully absorption-corrected magnitudes $M^{00}_{BVI}$
confirm the \citet{Pskovskii:67} effect of a correlation with the
$\Delta m_{15}$ decline rates (Figure~\ref{fig:other:dm15}) with
correlation slopes of 
$\alpha_{B}=0.619\pm0.076$,  
$\alpha_{V}=0.608\pm0.074$, 
$\alpha_{I}=0.438\pm0.076$ (Table~\ref{tab:other:dm15R}).
              
     The present slopes $\alpha_{BVI}$, which correct for the
magnitude differences of SNe\,Ia in E/S0 galaxies and spirals, 
lie between the smaller values of \citet{Parodi:etal:00} and the
larger values of \citet{Phillips:etal:99} and
\citet{Altavilla:etal:04}. It is shown that the reason for this is the
sensitivity of $\alpha_{BVI}$ on the adopted slopes of the
color-$\Delta m_{15}$ relation in
equations~(\ref{eq:color:mean:BV}$-$\ref{eq:color:mean:VI}). Steeper
slopes of these relations lead necessarily to larger values of
$\alpha_{BVI}$. Incorrect slopes of the
color-$\Delta m_{15}$ relation enter the color excesses and propagate
strongly into the absorption and hence into the $M^{00}$
magnitudes. They cancel, however, if the latter are normalized to a
common value of $\Delta m_{15}$ ($M^{00}_{15}$).
  
       (8) The absolute magnitudes, $M^{00}_{BVI\max}$ show some
dependence on the $(B\!-\!V)^{00}_{\max}$ and $(V\!-\!I)^{00}_{\max}$
colors (Figure~\ref{fig:other:color}), as suggested by
\citet{Tripp:Branch:99} and \citet{Parodi:etal:00}.  
Therefore the adopted magnitudes $M^{\rm corr}_{BV}$ of normal SNe\,Ia
have been reduced to common values of $\Delta m_{15}=1.1$ {\em and\/}
$(B\!-\!V)^{00}_{\max}=-0.024$ by equation~(\ref{eq:other:3fitd}) and
Table~\ref{tab:coeff3fit}. 
This slightly reduces the dispersion about the Hubble line without
changing the {\em mean\/} magnitude of the sample.

     (9) The present sample of 124 SNe\,Ia contains only four
spectroscopically distinct SNe-91T, with SN\,1991T as the
prototype. Their apparent rarity must be real because of their
pronounced overluminosity.  The overluminosity amounts to $\Delta
M_{B}=-0.40\pm0.08$ on average if one assumes them to have the same
color at maximum $(B\!-\!V)^{00}_{\max}$ as normal SNe\,Ia
(Table~\ref{tab:SNeIa-T}). The mean overluminosity is reduced to
$-0.32\pm0.08$ if one allows in addition for their below-average
decline rates $\Delta m_{15}$. 

     The new class of SNe-99aa \citep{Li:etal:01b}, named after
SN\,1999aa, comprises seven objects in Table~\ref{tab:SN1}a (marked
with $\dagger$). They are spectroscopically between normal SNe\,Ia and
SNe-91T, but are difficult to recognize because their spectroscopic
peculiarity is confined to the pre-maximum phase. Their true frequency
in a distance-limited sample may therefore be as high as
$\sim\!14\%$. They are less overluminous than SNe-91T, i.e.\
$\Delta M_{B}=-0.30\pm0.10$, or $\Delta M_{B}=-0.19\pm0.10$ if reduced
to $\Delta m_{15}=1.1$, as compared to normal SNe\,Ia.
-- SNe-99aa have very slow decline rates ($<\!\Delta
m_{15}\!>=0.92$) and occur correspondingly in late-type galaxies. The
mean coded type of their host galaxies is $<\!T\!>=3.7\pm0.5$
($\sim$Sbc) as compared to $<\!T\!>=3.5\pm1.0$ for SNe-91T and
$<\!T\!>=1.8$ ($\sim$Sab) for normal SNe\,Ia. 
   
     The most luminous SN in the sample is probably the very peculiar
SN\,2000cx which has -- with plausible absorption corrections --
$M^{00}_{BV}\approx-20.3$.  
 
    (10) Seven SNe-91bg, whose eponymous star is SN\,1991bg, have
sufficient photometry to be included in Table~\ref{tab:SN1}c. They are
very red and spectroscopically peculiar. In accordance with their
exceptionally fast decline rates ($\Delta m_{15}\ge1.9$) they lie
preferentially in early-type galaxies ($<\!T\!>=-0.6$). This suggests
little absorption in their host galaxies; zero absorption is assumed
here to derive their intrinsic photometric properties
(Table~\ref{tab:SNeIa-bg}). The SNe-91bg emerge as a surprisingly
homogeneous and separate class with $B$ magnitudes at maximum about
$2\mag$ below that of normal SNe\,Ia. The luminosity of  SN\,1986G
seems to lie between SNe\,Ia and SNe-91bg (see
Table~\ref{tab:SNeIa-bg}). 

     (11) The Hubble diagrams in $B$, $V$, and $I$ for all 111 normal
SNe\,Ia in Table~\ref{tab:SN1}a are shown in Figure~\ref{fig:hubble} using
fully corrected apparent magnitudes $m^{\rm corr}_{BVI}$. Solutions
for the intercepts $C_{\lambda}$ of the Hubble line are given for
various subsets with different restrictions in Table~\ref{tab:hubble}.
The fiducial sample of 62 (58) SNe\,Ia with $3000<v<20\,000\kms$ is
taken as the most reliable sample (solution~3 in
Table~\ref{tab:hubble}). The resulting values of 
$C_{\lambda}$ inserted in equation~(\ref{eq:hubble:04}) and $H_{0}=60$
give 
$<\!M_{B}^{\rm corr}\!>=-19.57\pm0.02$,  
$<\!M_{V}^{\rm corr}\!>=-19.55\pm0.02$, and   
$<\!M_{I}^{\rm corr}\!>=-19.29\pm0.02$.   
The dispersion of the $m_{BVI}^{\rm corr}$ magnitudes (corrected for
Galactic and internal absorption and normalized to $\Delta
m_{15}=1.1$) about the Hubble line is  $\sigma_{BVI}=0.14\mag$
(solution~3 in Table~\ref{tab:hubble}). It increases for distances
shorter than $3000\kms$ due to peculiar motions (solutions~5 and
6). Much of the dispersion beyond $3000\kms$ is caused by random
errors of the host galaxy absorption, as seen by the SNe\,Ia with
minimum absorption in Table~\ref{tab:color:minred} which give
$\sigma_{I}=0.10\mag$, if restricted in addition to $v>5000\kms$.
This is for normal SNe\,Ia, after they are normalized to common values
of $\Delta m_{15}$ and $(B\!-\!V)^{00}_{\max}$, an {\em upper\/} limit
to their intrinsic luminosity dispersion in $I$ in view of remaining
observational magnitude errors. The intrinsic dispersion in $B$ and
$V$ may be somewhat larger. 

     In spite of the noticeable scatter introduced by the corrections
for host galaxy absorption, the {\em mean\/} of the corrections must
be nearly correct. This is demonstrated by solutions~3 and 6-8 in
Table~\ref{tab:hubble}, where different cuts of the accepted maximum
value of the total color excess $E(B\!-\!V)_{\rm t}$ lead to
statistically insignificant variations of the intercept $C_{\lambda}$
of only $\pm0.004$ (corresponding to $\pm0.02\mag$).  

     The combination of the absolute magnitude $M^{\rm corr}_{V}$ from
above with the individual apparent magnitudes $m^{\rm corr}_{V}$
yields the absorption-free luminosity distances of 111 parent galaxies
of normal SNe\,Ia (Table~\ref{tab:SN3}a, column~11). The distances are
particularly valuable for nearby galaxies because they are insensitive
to peculiar motions. If the value $H_{0}=60$ is to be replaced by any
other value, the listed distance moduli need only a bulk correction by
an additive term.

     Had we adopted a flat Universe with $\Omega_{\rm M}=1$ the value
of $C_{\lambda}$ in solution~3 of Table~\ref{tab:hubble} would change
by only $-0.009$. This would decrease $H_{0}$ by merely $2\%$. The
insensitivity of $H_{0}$ to the cosmological model is due to the
velocity cut at $v_{\rm CMB}<20\,000\kms$ of solution~3.

     (12) The main goal of the present paper is to determine an
up-to-date Hubble diagram using a large and unbiased sample of
Branch-normal SNe\,Ia, whose magnitudes are corrected for Galactic and
host galaxy absorption and normalized to standard values of the
decline rate $\Delta m_{15}$ and of the color
$(B\!-\!V)^{00}_{\max}$. Regardless whether one considers the 
unrestricted sample of non-local or restricted samples with only a
minimum of extinction one finds remarkably robust solutions for the
intercept $C_{\lambda}$, which through equation~(\ref{eq:hubble:04})
is decisive for the determination of the value of $H_{0}$. The
remaining uncertainty of $C_{\lambda}$ of $\sim\!0.004$ will carry an
error into the derived value of $H_{0}$ of only $1\%$. The consequence
is that the accuracy of the derived large-scale value of $H_{0}$ will
depend almost exclusively on the goodness of the luminosity
calibration that we can achieve in a forthcoming paper by means of the
SNe\,Ia in host galaxies with known Cepheids.

\acknowledgments
We are greatly indebted to Dr.~G.~Altavilla for kindly providing his
unpublished template-fitted $V_{\max}$ magnitudes for a large sample
of SNe\,Ia. We are equally indebted to Dr.~S.~Jha for his generosity
to make his unpublished tail colors $(B\!-\!V)_{35}$ available to
us. B.\,R. thanks an anonymous donator for financial support, while
G.\,A.\,T. thanks the Swiss National Science Foundation for support over
many decades. A.\,S. is indebted to the Carnegie Institution for
ongoing post-retirement facilities support.



\clearpage
\onecolumn
 \setlength\oddsidemargin   {-1.45cm}%
 \setlength\evensidemargin  {-1.45cm}%
\setcounter{table}{0}
\def\baselinestretch{1.0}
\begin{deluxetable}{llcrrrrrcrrrrl}
\tablewidth{540pt}
\tabletypesize{\scriptsize}
\tablecaption{Observed parameters of SNe\,Ia.\label{tab:SN1}}
\tablehead{
 \colhead{SN} &
 \colhead{Galaxy} &
 \colhead{T} &
 \colhead{$v_{\rm hel}$}&
 \colhead{$\!\!v_{\rm CMB/220}\!\!$}&
 \colhead{$B_{\max}$} &
 \colhead{$V_{\max}$} &
 \colhead{$I_{\max}$} &
 \colhead{$\!\Delta m_{15}\!$} &
 \colhead{$E(B\mbox{-}V)$}&
 \colhead{$E(B\mbox{-}V)$} &
 \colhead{$E(B\mbox{-}V)$} &
 \colhead{$E(B\mbox{-}V)$} &
 \colhead{Ref.} \\
 \colhead{}  & \colhead{}  &
 \colhead{}  & \colhead{}  &
 \colhead{}  & \colhead{}  &
 \colhead{}  & \colhead{}  &
 \colhead{}  & \colhead{Gal} &
 \colhead{max} & \colhead{max35} &
 \colhead{$V\!-\!I$} & \colhead{} \\
 \colhead{(1)}  & \colhead{(2)}  &
 \colhead{(3)}  & \colhead{(4)}  &
 \colhead{(5)}  & \colhead{(6)}  &
 \colhead{(7)}  & \colhead{(8)}  &
 \colhead{(9)}  & \colhead{(10)} &
 \colhead{(11)} & \colhead{(12)} &
 \colhead{(13)} & \colhead{(14)}
}
\startdata
\multicolumn{14}{c}{\footnotesize (a) normal SNe\,Ia}\\
\noalign{\smallskip}
\tableline
1937C              & IC 4182$^{\ast}$   & 5       & 321     &  330$^{\rm v}$ &  8.80   &  8.82   & \nodata & 0.85    & 0.014 &  0.001 & -0.045 &\nodatr &  A   \\  
1960F              & NGC 4496A$^{\ast}$ & 5       & 1730    & 1179$^{\rm V}$ & 11.60   & 11.51   & \nodata & 0.87    & 0.025 &  0.099 &\nodatr &\nodatr &  Sa  \\  
1972E              & NGC 5253$^{\ast}$  & 5       & 404     &  167$^{\rm v}$ &  8.49   &  8.49   &  8.80   & 1.05    & 0.056 & -0.030 & -0.052 & -0.069 &  Sa  \\  
1972J              & NGC 7634           & -1      & 3225    & 2855           & 14.80   & 14.56   & \nodata & 1.45    & 0.046 &  0.202 & -0.040 &\nodatr &  A   \\  
1974G              & NGC 4414$^{\ast}$  & 5       & 716     &  655$^{\rm v}$ & 12.48   & 12.30   & \nodata & 1.11    & 0.019 &  0.184 &  0.137 &\nodatr &  S   \\  
1980N              & NGC 1316           & 1       & 1760    & 1338$^{\rm F}$ & 12.49   & 12.42   & 12.70   & 1.30    & 0.021 &  0.064 &  0.039 & -0.045 &  HA  \\  
1981B              & NGC 4536$^{\ast}$  & 4       & 1808    &(1179$^{\rm V}$)& 12.04   & 11.98   & \nodata & 1.13    & 0.018 &  0.064 &  0.009 &\nodatr &  A   \\  
1981D              & NGC 1316           & 1       & 1760    & 1338$^{\rm F}$ & 12.59   & 12.40   & \nodata & \nodata & 0.021 &  0.242 &\nodatr &\nodatr &  P   \\
1982B              & NGC 2268           & 4       & 2222    & 2610$^{\rm v}$ & 13.65   & 13.40   & \nodata & 0.94    & 0.064 &  0.217 &  0.041 &\nodatr &  A   \\  
1983G              & NGC 4753           & -1      & 1239    & 1497$^{\rm v}$ & 13.00   & 12.80   & \nodata & 1.30    & 0.034 &  0.180 &  0.281 &\nodatr &  A   \\  
1984A              & NGC 4419           & 1       & -261    & 1179$^{\rm V}$ & 12.50   & 12.30   & \nodata & \nodata & 0.033 &  0.240 &\nodatr &\nodatr &  H   \\  
1989B              & NGC 3627$^{\ast}$  & 3       & 727     &  549$^{\rm v}$ & 12.34   & 11.99   & 11.70   & 1.31    & 0.032 &  0.332 &  0.298 &  0.302 &  W   \\  
1990N              & NGC 4639$^{\ast}$  & 3       & 1010    & 1179$^{\rm V}$ & 12.75   & 12.73   & 12.95   & 1.05    & 0.026 &  0.020 &  0.070 &  0.012 &  HA  \\  
1990O              & MCG 3-44-03        & 1       & 9193    & 9175           & 16.59   & 16.51   & 16.80   & 0.94    & 0.093 &  0.018 & -0.001 & -0.076 &  HA  \\  
1990T              & PGC 63925          & 1       & 11992   & 11893          & 17.27   & 17.35   & 17.36   & 1.13    & 0.053 & -0.111 &  0.055 &  0.114 &  HA  \\  
1990Y              & Anon 0335-33       & -3      & 11702   & 11603          & 17.69   & 17.42   & 17.57   & 1.13    & 0.008 &  0.284 &  0.205 &  0.063 &  HA  \\  
1990af             & Anon 2135-62       & -1      & 15080   & 14966          & 17.91   & 17.84   & \nodata & 1.59    & 0.035 &  0.036 & -0.022 &\nodatr &  H   \\  
1991S              & UGC 5691           & 3       & 16489   & 16807          & 17.78   & 17.75   & 17.99   & 1.00    & 0.026 &  0.032 &  0.033 &  0.004 &  HA  \\  
1991U              & IC 4232            & 4       & 9426    & 9724           & 16.67   & 16.61   & 16.62   & 1.11    & 0.062 &  0.021 &  0.056 &  0.108 &  HA  \\  
1991ag             & IC 4919            & 3       & 4264    & 4161           & 14.67   & 14.56   & 14.83   & 0.87    & 0.062 &  0.082 & -0.012 & -0.032 &  HA  \\  
1992A              & NGC 1380           & 0       & 1877    & 1338$^{\rm F}$ & 12.56   & 12.55   & 12.80   & 1.47    & 0.018 & -0.001 & -0.032 & -0.040 &  HA  \\  
1992J              & Anon 1009-26       & -2      & 13491   & 13828          & 17.88   & 17.76   & 17.88   & 1.69    & 0.057 &  0.060 &  0.032 & -0.011 &  HA  \\  
1992P              & IC 3690            & 1       & 7615    & 7941           & 16.14   & 16.16   & 16.40   & 1.05    & 0.021 & -0.015 &  0.014 &  0.003 &  HA  \\  
1992ae             & Anon 2128-61       & -3      & 22484   & 22366          & 18.62   & 18.54   & \nodata & 1.30    & 0.036 &  0.059 &  0.053 &\nodatr &  A   \\  
1992ag             & ESO 508-G67        & 5       & 7795    & 8095           & 16.64   & 16.47   & 16.50   & 1.20    & 0.097 &  0.092 &  0.359 &  0.059 &  HA  \\  
1992al             & ESO 234-G69        & 3       & 4197    & 4048           & 14.59   & 14.65   & 14.93   & 1.09    & 0.034 & -0.070 & -0.015 & -0.036 &  HA  \\  
1992aq             & Anon 2304-37       & 1       & 30279   & 30014          & 19.42   & 19.39   & 19.48   & 1.69    & 0.012 &  0.015 &  0.026 &  0.044 &  HA  \\  
1992au             & Anon 0010-49       & -3      & 18287   & 18092          & 18.17   & 18.13   & 18.51   & 1.69    & 0.017 &  0.020 & -0.014 & -0.141 &  HA  \\  
1992bc             & ESO 300-G9         & 2       & 5996    & 5876           & 15.13   & 15.25   & 15.56   & 0.90    & 0.022 & -0.109 & -0.108 & -0.027 &  HA  \\  
1992bg             & Anon 0741-62       & 1       & 10793   & 10936          & 17.39   & 17.30   & 17.32   & 1.15    & 0.185 & -0.074 & -0.015 & -0.002 &  HA  \\  
1992bh             & Anon 0459-58       & 4       & 13491   & 13519          & 17.68   & 17.63   & 17.75   & 1.13    & 0.022 &  0.050 &  0.061 &  0.070 &  HA  \\  
1992bk             & ESO 156-G8         & -3      & 17284   & 17237          & 18.07   & 18.14   & 18.21   & 1.67    & 0.015 & -0.087 & -0.018 &  0.056 &  HA  \\  
1992bl             & ESO 291-G11        & 0       & 12891   & 12661          & 17.34   & 17.37   & 17.59   & 1.56    & 0.011 & -0.038 & -0.033 & -0.024 &  HA  \\  
1992bo             & ESO 352-G57        & -2      & 5396    & 5164           & 15.86   & 15.84   & 15.95   & 1.69    & 0.027 & -0.010 & -0.030 &  0.020 &  HA  \\  
1992bp             & Anon 0336-18       & -2      & 23684   & 23557          & 18.53   & 18.53   & 18.68   & 1.52    & 0.069 & -0.064 &  0.020 & -0.024 &  HA  \\  
1992br             & Anon 0145-56       & -3      & 26382   & 26259          & 19.34   & 19.27   & \nodata & 1.69    & 0.026 &  0.041 & -0.020 &\nodatr &  H   \\  
1992bs             & Anon 0329-37       & 3       & 18887   & 18787          & 18.33   & 18.33   & \nodata & 1.15    & 0.011 &  0.010 &  0.017 &\nodatr &  A   \\  
1993B              & Anon 1034-34       & 3       & 20686   & 21011          & 18.71   & 18.65   & 18.71   & 1.30    & 0.079 & -0.004 &  0.122 &  0.045 &  HA  \\  
1993H              & ESO 445-G66        & 2       & 7257    & 7522           & 16.99   & 16.72   & 16.60   & 1.69    & 0.060 &  0.207 &  0.022 &  0.137 &  HA  \\  
1993L              & IC 5270            & 4       & 1926    & 1854$^{\rm v}$ & 13.40   & 13.22   & \nodata & 1.47    & 0.014 &  0.173 &  0.212 &\nodatr &  A   \\  
1993O              & Anon 1331-33       & -2      & 15589   & 15867          & 17.79   & 17.86   & 17.99   & 1.26    & 0.053 & -0.107 & -0.012 &  0.026 &  HA  \\  
1993ac             & PGC 17787          & -3      & 14690   & 14674          & 18.42   & 18.25   & 18.12   & 1.25    & 0.163 &  0.024 & -0.004 &  0.101 &  RPA \\  
1993ae             & IC 126             & -3      & 5712    & 5410           & 15.42   & 15.52   & 15.65   & 1.47    & 0.038 & -0.131 & -0.058 &  0.019 &  RPA \\  
1993ag             & Anon 1003-35       & -2      & 14700   & 15013          & 18.30   & 18.12   & 18.22   & 1.30    & 0.112 &  0.082 &  0.055 & -0.007 &  HA  \\  
1993ah             & ESO 471-G27        & -1      & 8803    & 8504           & 16.32   & 16.36   & 16.65   & 1.45    & 0.020 & -0.052 &  0.048 & -0.065 &  HA  \\  
1994D              & NGC 4526           & -1      & 448     & 1179$^{\rm V}$ & 11.86   & 11.94   & 12.11   & 1.31    & 0.022 & -0.088 & -0.054 &  0.022 &  HA  \\  
1994M              & NGC 4493           & -3      & 6943    & 7289           & 16.35   & 16.30   & 16.40   & 1.45    & 0.023 &  0.035 &  0.083 &  0.052 &  RPA \\  
1994Q              & PGC 59076          & -1      & 8863    & 8861           & 16.44   & 16.36   & 16.63   & 0.90    & 0.017 &  0.096 &  0.007 &  0.002 &  RPA \\  
1994S              & NGC 4495           & 3       & 4550    & 4831           & 14.79   & 14.83   & 15.14   & 1.02    & 0.021 & -0.034 & -0.055 & -0.038 &  R   \\  
1994T              & PGC 46640          & 1       & 10390   & 10703          & 17.35   & 17.23   & 17.38   & 1.45    & 0.029 &  0.099 &  0.028 &  0.016 &  RPA \\  
1994ae             & NGC 3370           & 5       & 1279    & 1575$^{\rm v}$ & 13.15   & 13.10   & 13.40   & 0.87    & 0.030 &  0.054 &  0.070 & -0.025 &  RPA \\  
1995D              & NGC 2962           & 0       & 1966    & 2129$^{\rm v}$ & 13.44   & 13.40   & 13.70   & 1.03    & 0.058 &  0.009 &  0.028 & -0.063 &  RPA \\  
1995E              & NGC 2441           & 3       & 3470    & 3496           & 16.82   & 16.10   & \nodata & 1.19    & 0.027 &  0.712 &  0.582 &\nodatr &  RA  \\  
1995ak             & IC 1844            & 1       & 6811    & 6589           & 16.09   & 16.10   & 16.21   & 1.45    & 0.038 & -0.040 &  0.172 &  0.033 &  RPA \\  
1995al             & NGC 3021           & 4       & 1541    & 1851$^{\rm v}$ & 13.36   & 13.25   & 13.50   & 0.89    & 0.014 &  0.129 &  0.066 &  0.018 &  RPA \\  
1996C              & MCG 8-25-47        & 1       & 8094    & 8245           & 16.59   & 16.56   & 16.77   & 0.94    & 0.013 &  0.048 & -0.007 &  0.039 &  RPA \\  
1996X              & NGC 5061           & -3      & 2039    & 2120$^{\rm v}$ & 13.26   & 13.21   & 13.39   & 1.32    & 0.069 & -0.005 & -0.016 & -0.024 &  RPA \\  
1996Z              & NGC 2935           & 3       & 2277    & 2285$^{\rm v}$ & 14.61   & 14.25   & \nodata & 1.22    & 0.064 &  0.314 &  0.350 &\nodatr &  R   \\  
1996ab         & Anon 1521+28$\!\!\!\!$ & 5       & 37240   & 37370          & 19.54   & 19.42   & \nodata & 0.87    & 0.032 &  0.122 &  0.080 &\nodatr &  R   \\  
1996ai$^{1)}$      & NGC 5005           & 4       & 946     & 1298$^{\rm v}$ & 16.96   & 15.26   & \nodata & 0.85    & 0.014 &  1.721 &  1.501 &\nodatr &  RA  \\  
1996bk             & NGC 5308           & -1      & 2041    & 2462$^{\rm v}$ & 14.84   & 14.41   & \nodata & 1.69    & 0.018 &  0.409 &  0.207 &\nodatr &  RA  \\  
1996bl       & Anon 0036+11$\!\!\!\!$   & 5       & 10793   & 10447          & 17.08   & 16.98   & 17.03   & 1.11    & 0.092 &  0.031 &  0.036 &  0.059 &  RPA \\  
1996bo             & NGC 673            & 5       & 5182    & 4893           & 16.15   & 15.78   & \nodata & 1.30    & 0.077 &  0.307 &  0.206 &\nodatr &  RA  \\  
1996bv             & UGC 3432           & 5       & 4998    & 5016           & 15.77   & 15.54   & 15.56   & 0.84    & 0.105 &  0.160 &  0.098 &  0.092 &  RPA \\  
1997E              & NGC 2258           & -1      & 4001    & 3998           & 15.59   & 15.45   & 15.46   & 1.39    & 0.124 &  0.026 &  0.013 &  0.031 &  J   \\  
1997Y              & NGC 4675           & 3       & 4806    & 4964           & 15.28   & 15.30   & 15.36   & 1.25    & 0.017 & -0.020 &  0.024 &  0.100 &  J   \\  
1997bp             & NGC 4680           & 3       & 2492    & 2647$^{\rm v}$ & 14.12   & 13.92   & 14.09   & 0.97    & 0.044 &  0.185 &  0.286 &  0.036 &  AJ  \\  
1997bq             & NGC 3147           & 4       & 2820    & 2879           & 14.57   & 14.27   & 14.37   & 1.01    & 0.024 &  0.304 &  0.082 &  0.092 &  J   \\  
1997cw$^{\dagger}$ & NGC 105            & 2       & 5133    & 4782           & 16.00   & 15.52   & 15.33   & 1.02    & 0.073 &  0.434 &  0.335 &  0.233 &  J   \\  
1997dg             & anonymous          & 1       & 10193   & 9845           & 17.20   & 17.11   & 17.14   & 1.13    & 0.078 &  0.034 &\nodatr &  0.081 &  J   \\  
1997do             & UGC 3845           & 4       & 3034    & 3140           & 14.56   & 14.46   & 14.59   & 0.99    & 0.063 &  0.065 &\nodatr &  0.044 &  J   \\  
1997dt             & NGC 7448           & 4       & 2194    & 2356$^{\rm v}$ & 15.64   & 15.08   & 14.54   & 1.04    & 0.057 &  0.529 &\nodatr &  0.463 &  J   \\  
1998V              & NGC 6627           & 3       & 5268    & 5148           & 15.88   & 15.71   & 15.62   & 1.06    & 0.196 & -0.001 & -0.018 &  0.066 &  J   \\  
1998ab$^{\dagger}$ & NGC 4704           & 5       & 8134    & 8354           & 15.94   & 15.93   & 15.98   & 0.88    & 0.017 &  0.026 &  0.070 &  0.141 &  J   \\  
1998aq             & NGC 3982$^{\ast}$  & 3       & 1109    & 1514$^{\rm v}$ & 12.22   & 12.28   & \nodata & 1.12    & 0.014 & -0.051 & -0.053 &\nodatr &  Sa  \\  
1998bu             & NGC 3368$^{\ast}$  & 2       & 897     & 810$^{\rm v}$  & 12.22   & 11.86   & 11.65   & 1.15    & 0.025 &  0.356 &  0.209 &  0.272 &  A   \\  
1998dh             & NGC 7541           & 5       & 2678    & 2766$^{\rm v}$ & 14.24   & 13.98   & 14.04   & 1.23    & 0.068 &  0.210 &  0.090 &  0.061 &  J   \\  
1998dk             & UGC 139            & 5       & 3963    & 3609           & 14.93   & 14.74   & 14.81   & 1.05    & 0.044 &  0.172 &  0.127 &  0.091 &  J   \\  
1998dm             & MCG-01-4-44$\!\!$  & 4       & 1968    & 1976$^{\rm v}$ & 14.70   & 14.48   & 14.28   & 1.07    & 0.044 &  0.201 &  0.259 &  0.258 &  J   \\  
1998dx             & UGC 11149          & -3      & 14990   & 14895          & 17.71   & 17.74   & 17.86   & 1.55    & 0.041 & -0.068 &\nodatr &  0.015 &  J   \\  
1998ec             & UGC 3576           & 3       & 5966    & 6032           & 16.44   & 16.21   & 16.22   & 1.08    & 0.085 &  0.169 &\nodatr &  0.092 &  J   \\  
1998ef             & UGC 646            & 3       & 5319    & 5020           & 15.21   & 15.18   & 15.28   & 0.97    & 0.073 & -0.014 &\nodatr &  0.056 &  J   \\  
1998eg             & UGC 12133          & 5       & 7423    & 7056           & 16.62   & 16.50   & 16.49   & 1.15    & 0.123 &  0.018 &\nodatr &  0.067 &  J   \\  
1998es$^{\dagger}$ & NGC 632            & 2       & 3168    & 2868           & 13.99   & 13.87   & 14.11   & 0.87    & 0.032 &  0.122 &  0.031 &  0.011 &  J   \\  
1999X           & CGCG 180-22$\!\!\!\!$ & \nodata & 7503    & 7720           & 16.45   & 16.29   & 16.31   & 1.11    & 0.032 &  0.151 &\nodatr &  0.126 &  J   \\  
1999aa$^{\dagger}$ & NGC 2595           & 5       & 4330    & 4572           & 14.92   & 14.88   & 15.24   & 0.85    & 0.040 &  0.035 & -0.033 & -0.069 &  AJ  \\  
1999ac$^{\dagger}$ & NGC 6063           & 4       & 2848    & 2944           & 14.34   & 14.28   & 14.32   & 1.00    & 0.046 &  0.042 &\nodatr &  0.113 &  J   \\  
1999aw         & Anon 1101-06$\!\!\!\!$ & \nodata & 11992   & 12363          & 16.86   & 16.84   & 17.24:  & 0.81    & 0.032 &  0.025 & -0.076 & -0.083 &  SlA \\  
1999cc             & NGC 6038           & 5       & 9392    & 9452           & 16.85   & 16.82   & 17.03   & 1.46    & 0.023 &  0.014 &\nodatr & -0.018 &  J   \\  
1999cl             & NGC 4501           & 3       & 2281    & 1179$^{\rm V}$ & 15.11   & 13.89   & 13.08   & 1.19    & 0.038 &  1.201 &\nodatr &  0.634 &  J   \\  
1999dk             & UGC 1087           & 5       & 4485    & 4181           & 15.04   & 14.95   & 15.22   & 1.28    & 0.054 &  0.051 &  0.084 & -0.064 &  KA  \\  
1999dq$^{\dagger}$ & NGC 976            & 5       & 4295    & 4060           & 14.88   & 14.68   & 14.76   & 0.88    & 0.110 &  0.123 &  0.119 &  0.047 &  J   \\  
1999ee             & IC 5179            & 4       & 3422    & 3163           & 14.93   & 14.61   & 14.66   & 0.92    & 0.020 &  0.332 &  0.237 &  0.135 &  St  \\  
1999ef             & UGC 607            & 5       & 11733   & 11402          & 17.52   & 17.39   & 17.69   & 1.06    & 0.087 &  0.068 &\nodatr & -0.089 &  J   \\  
1999ej             & NGC 495            & 0       & 4114    & 3831           & 15.65   & 15.60   & 15.66   & 1.41    & 0.071 & -0.011 &\nodatr &  0.041 &  J   \\  
1999ek             & UGC 3329           & 4       & 5253    & 5278           & 17.97   & 17.19   & 16.54   & 1.13    & 0.561 &  0.241 &\nodatr &  0.113 &  J   \\  
1999gd             & NGC 2623           & \nodata & 5535    & 5775           & 17.02   & 16.61   & 16.27   & 1.14    & 0.041 &  0.391 &\nodatr &  0.342 &  J   \\  
1999gh             & NGC 2986           & -3      & 2302    & 2314$^{\rm v}$ & 14.46   & 14.27   & 14.25   & 1.69    & 0.058 &  0.129 &  0.046 &  0.076 &  J   \\  
1999gp$^{\dagger}$ & UGC 1993           & 3       & 8018    & 7806           & 16.25   & 16.10   & 16.43   & 0.94    & 0.056 &  0.125 &  0.027 & -0.071 &  AJ  \\  
2000B              & NGC 2320           & -2      & 5944    & 6020           & 15.94   & 15.77   & 15.94   & 1.46    & 0.068 &  0.109 &  0.028 & -0.030 &  J   \\  
2000E              & NGC 6951           & 3       & 1424    & 1824$^{\rm v}$ & 14.31   & 13.83   & 13.49   & 0.94    & 0.366 &  0.145 &  0.167 &  0.096 &  VA  \\  
2000bk             & NGC 4520           & 0       & 7628    & 7974           & 16.98   & 16.95   & 16.81   & 1.69    & 0.025 &  0.002 &  0.116 &  0.178 &  KA  \\  
2000ce             & UGC 4195           & 3       & 4888    & 4940           & 17.31   & 16.66   & 16.22   & 1.00    & 0.057 &  0.621 &  0.445 &  0.404 &  AJ  \\  
2000cf             & MCG 11-19-25$\!\!\!\!$       & \nodata & 10793 & 10805  & 17.15   &  7.10   & 17.31   & 1.27    & 0.032 &  0.034 &\nodatr & -0.008 &  J   \\  
2000cn             & UGC 11064          & 5       & 7043    & 6969           & 16.82   & 16.63   & 16.71   & 1.58    & 0.057 &  0.135 & -0.061 &  0.024 &  J   \\  
2000dk             & NGC 382            & -3      & 5229    & 4932           & 15.63   & 15.56   & 15.74   & 1.57    & 0.070 &  0.002 & -0.062 & -0.048 &  J   \\  
2000fa             & UGC 3770           & 5       & 6378    & 6533           & 15.99   & 15.95   & 16.12   & 1.00    & 0.069 & -0.001 &  0.044 &  0.013 &  J   \\  
2001V              & NGC 3987           & 3       & 4502    & 4810           & 14.64   & 14.60   & 14.84   & 0.99    & 0.020 &  0.048 &  0.036 &  0.010 &  ViM \\
2001el             & NGC 1448           & 5       & 1164    & 1030$^{\rm v}$ & 12.81   & 12.73   & 12.82   & 1.13    & 0.014 &  0.088 &  0.201 &  0.095 &  K   \\  
2002bo             & NGC 3190           & 1       & 1271    & 1547$^{\rm v}$ & 14.04   & 13.58   & 13.52   & 1.13    & 0.025 &  0.457 &  0.311 &  0.180 &  B   \\  
2002er             & UGC 10743          & 3       & 2569    & 2804$^{\rm v}$ & 14.89   & 14.59   & 14.49   & 1.33    & 0.157 &  0.156 &  0.116 &  0.079 &  Pi   \\  
\tableline
\noalign{\smallskip}
\multicolumn{14}{c}{\footnotesize (b) SNe-91T}\\
\noalign{\smallskip}
\tableline
1991T              & NGC 4527$^{\ast}$  & 3       & 1736    &(1179$^{\rm V}$)& 11.69   & 11.50   & 11.62   & 0.94    & 0.022 &  0.199 & (0.064)$^{2)}$  &  0.088 & HA  \\  
1995ac             & Anon 2245-08       & 1       & 14990   & 14635          & 17.21   & 17.21   & 17.36   & 0.95    & 0.042 & -0.012 & (-0.011)&  0.052 & RPA \\  
1995bd             & UGC 3151           & 5       & 4377    & 4326           & 17.27   & 16.50   & \nodata & 0.88    & 0.498 &  0.305 & (0.109) &  \nodatr & RA  \\  
1997br             & ESO 576-40         & 5       & 2085    & 2193$^{\rm v}$ & 14.04   & 13.68   & 13.44   & 1.03    & 0.113 &  0.274 & (0.198) &  0.231 & AJ  \\  
\noalign{\smallskip}
2000cx             & NGC 524            & -1      & 2379    & 2454$^{\rm v}$ & 13.44   & 13.27   & 13.62   & 0.93    & 0.083 &   (0.118) & (-0.228)& -0.105 & AL  \\  
\tableline
\noalign{\smallskip}
\multicolumn{14}{c}{\footnotesize (c) SNe-91bg}\\
\noalign{\smallskip}
\tableline
1991bg             & NGC 4374           & -3      & 1060    & 1179$^{\rm V}$ & 14.75   & 13.96   & 13.51   & 1.94    & 0.041 & ---~~ &  (0.005)$^{2)}$   & ---~~ & HA  \\  
1992K              & ESO 269-G57$\!\!\!$& 3       & 3080    & 3334           & 16.31   & 15.42   & 15.16   & 1.94    & 0.101 & ---~~ & (-0.029)  & ---~~ & HA  \\  
1997cn             & NGC 5490           & -3      & 4855    & 5092           & 16.93   & 16.39   & 16.14   & 1.90    & 0.027 & ---~~ & \nodatr   & ---~~ & J   \\  
1998bp             & NGC 6495           & -3      & 3127    & 3048           & 15.73   & 15.25   & 15.01   & 1.83    & 0.076 & ---~~ &  (0.059)  & ---~~ & J   \\  
1998de             & NGC 252            & 2       & 4990    & 4671           & 17.56   & 16.80   & 16.48   & 1.94    & 0.058 & ---~~ &  (0.053)  & ---~~ & AJ  \\  
1999by             & NGC 2841$^{\ast}$  & 3       &  638    & 896$^{\rm v}$  & 13.66   & 13.15   & 12.91   & 1.90    & 0.016 & ---~~ &  (0.011)  & ---~~ & G   \\
1999da             & NGC 6411           & -3      & 3806    & 3748           & 16.90   & 16.16   & 15.85   & 1.94    & 0.058 & ---~~ &  (0.027)  & ---~~ & KA  \\
\noalign{\smallskip}
1986G              & NGC 5128           & -1      & 547     & 317$^{\rm v}$  & 12.48   & 11.44   & \nodata & 1.69    & 0.115 &  (0.922) & (0.489) & \nodatr & A   \\  
\enddata
\tablerefs{[A] \citet{Altavilla:etal:04},
          [B] \citet{Benetti:etal:04},
          [G] \citet{Garnavich:etal:04},
          [H] \citet{Hamuy:etal:96b,Hamuy:etal:96c,Hamuy:etal:96d},
          [J] \citet{Jha:02},
          [K] \citet{Krisciunas:etal:01,Krisciunas:etal:03}, 
          [L] \citet{Li:etal:01c},
          [M] \citet{Mandel:etal:01}
          [P] \citet{Parodi:etal:00},
         [Pi] \citet{Pignata:etal:04},
          [R] \citet{Riess:etal:99},
         [Sa] \citet{Saha:etal:96,Saha:etal:01b},
          [S] \citet{Schaefer:98},
         [Sl] \citet{Strolger:etal:02},
         [St] \citet{Stritzinger:etal:02},
          [V] \citet{Valentini:etal:03},
         [Vi] \citet{Vinko:etal:03},
          [W] \citet{Wells:etal:94}.}
\tablenotetext{\dagger}{99aa-like SNe\,Ia. See \S~\ref{sec:SNpec:T}.}
\tablenotetext{\ast}{Host galaxies for which a Cepheid
  distance is available.} 
\tablenotetext{1)}{The SN suffers exceptionally large
  absorption in the host galaxy \citep{Bottari:etal:96}; it is not
  used in the following.}
\tablenotetext{2)}{The values of $E(B\!-\!V)_{\max35}$ for SNe-91T,
  SNe-91bc, and for SN\,2000cx and 1986G are not used in the
  following. They are calculated on the assumption that their
  intrinsic color $(B\!-\!V)^{00}_{35}$ is given by
  equation~(\ref{eq:color:mean:BV35}) as for normal SNe\,Ia.} 
\end{deluxetable}

\clearpage
\onecolumn
 \setlength\oddsidemargin   {-0.20cm}%
 \setlength\evensidemargin  {-0.20cm}%
\setcounter{table}{2}
\def\baselinestretch{1.1}
\begin{deluxetable}{lrrrrrrrrrc}
\tablewidth{0pt}
\tabletypesize{\scriptsize}
\tablecaption{Derived parameters of SNe\,Ia. \label{tab:SN3}}
\tablehead{
 \colhead{SN} &
 \colhead{$E(B\mbox{-}V)_{\rm h}$}&
 \colhead{$(B\mbox{-}V)^{00}$} &
 \colhead{$(V\mbox{-}I)^{00}$} &
 \colhead{$m^{00}_{B}$} &
 \colhead{$m^{00}_{V}$} &
 \colhead{$m^{00}_{I}$} &
 \colhead{$m^{\rm corr}_{B}$} &
 \colhead{$m^{\rm corr}_{V}$} &
 \colhead{$m^{\rm corr}_{I}$} &
 \colhead{$(m-M)^{00}_{\rm lum}$} \\
 \colhead{(1)}  & \colhead{(2)}  &
 \colhead{(3)}  & \colhead{(4)}  &
 \colhead{(5)}  & \colhead{(6)}  &
 \colhead{(7)}  & \colhead{(8)}  &
 \colhead{(9)}  & \colhead{(10)}  &
 \colhead{(11)} 
}
\startdata
\multicolumn{11}{c}{\footnotesize (a) normal SNe\,Ia}\\
\noalign{\smallskip}
\hline
1937C  & -0.022 & -0.012 &\nodatr &  8.823 &  8.835 &\nodatr &  8.967 &  8.991 &\nodatr &  28.54  \\
1960F  &  0.099 & -0.034 &\nodatr & 11.137 & 11.171 &\nodatr & 11.284 & 11.308 &\nodatr &  30.86  \\
1972E  & -0.050 & -0.006 & -0.316 &  8.444 &  8.450 &  8.767 &  8.462 &  8.486 &  8.774 &  28.04  \\
1972J  &  0.081 &  0.113 &\nodatr & 14.316 & 14.203 &\nodatr & 14.007 & 14.031 &\nodatr &  33.58  \\
1974G  &  0.161 &  0.000 &\nodatr & 11.816 & 11.815 &\nodatr & 11.793 & 11.817 &\nodatr &  31.37  \\
1980N  &  0.019 &  0.030 & -0.332 & 12.334 & 12.304 & 12.637 & 12.175 & 12.199 & 12.504 &  31.75  \\
1981B  &  0.037 &  0.005 &\nodatr & 11.832 & 11.827 &\nodatr & 11.794 & 11.818 &\nodatr &  31.37  \\
1981D  &  0.193 & -0.024 &\nodatr & 11.801 & 11.825 &\nodatr & 11.801 & 11.825 &\nodatr &  31.38  \\
1982B  &  0.129 &  0.057 &\nodatr & 12.917 & 12.860 &\nodatr & 12.958 & 12.982 &\nodatr &  32.53  \\
1983G  &  0.231 & -0.065 &\nodatr & 12.018 & 12.083 &\nodatr & 11.924 & 11.948 &\nodatr &  31.50  \\
1984A  &  0.190 & -0.024 &\nodatr & 11.669 & 11.693 &\nodatr & 11.669 & 11.693 &\nodatr &  31.24  \\
1989B  &  0.311 &  0.007 & -0.162 & 11.075 & 11.068 & 11.223 & 10.925 & 10.949 & 11.105 &  30.50  \\
1990N  &  0.034 & -0.040 & -0.298 & 12.520 & 12.560 & 12.858 & 12.562 & 12.586 & 12.893 &  32.14  \\
1990O  & -0.020 &  0.007 & -0.385 & 16.281 & 16.274 & 16.659 & 16.357 & 16.381 & 16.704 &  35.93  \\
1990T  &  0.019 & -0.152 & -0.104 & 16.983 & 17.135 & 17.239 & 17.053 & 17.077 & 17.332 &  36.63  \\
1990Y  &  0.184 &  0.078 & -0.403 & 16.985 & 16.907 & 17.307 & 16.897 & 16.921 & 17.210 &  36.47  \\
1990af &  0.007 &  0.028 &\nodatr & 17.739 & 17.712 &\nodatr & 17.404 & 17.428 &\nodatr &  36.98  \\
1991S  &  0.023 & -0.019 & -0.304 & 17.590 & 17.609 & 17.912 & 17.647 & 17.671 & 17.952 &  37.22  \\
1991U  &  0.062 & -0.064 & -0.172 & 16.191 & 16.255 & 16.425 & 16.212 & 16.236 & 16.454 &  35.79  \\
1991ag &  0.013 &  0.035 & -0.367 & 14.369 & 14.334 & 14.701 & 14.469 & 14.493 & 14.753 &  34.04  \\
1992A  & -0.025 &  0.017 & -0.241 & 12.576 & 12.559 & 12.801 & 12.321 & 12.345 & 12.605 &  31.90  \\
1992J  &  0.027 &  0.036 & -0.230 & 17.547 & 17.511 & 17.741 & 17.145 & 17.169 & 17.432 &  36.72  \\
1992P  &  0.001 & -0.042 & -0.268 & 16.052 & 16.093 & 16.361 & 16.094 & 16.118 & 16.398 &  35.67  \\
1992ae &  0.056 & -0.012 &\nodatr & 18.268 & 18.280 &\nodatr & 18.138 & 18.162 &\nodatr &  37.71  \\
1992ag &  0.170 & -0.097 & -0.380 & 15.623 & 15.719 & 16.096 & 15.612 & 15.636 & 16.112 & (35.19) \\
1992al & -0.040 & -0.054 & -0.271 & 14.598 & 14.651 & 14.923 & 14.624 & 14.648 & 14.952 &  34.20  \\
1992aq &  0.028 & -0.010 & -0.143 & 19.267 & 19.277 & 19.420 & 18.896 & 18.920 & 19.150 &  38.47  \\
1992au & -0.045 &  0.068 & -0.343 & 18.265 & 18.197 & 18.540 & 17.840 & 17.864 & 18.205 &  37.41  \\
1992bc & -0.081 & -0.061 & -0.231 & 15.337 & 15.398 & 15.630 & 15.485 & 15.509 & 15.748 &  35.06  \\
1992bg & -0.030 & -0.065 & -0.221 & 16.742 & 16.807 & 17.028 & 16.740 & 16.764 & 17.040 &  36.31  \\
1992bh &  0.060 & -0.032 & -0.228 & 17.369 & 17.402 & 17.629 & 17.357 & 17.381 & 17.623 &  36.93  \\
1992bk & -0.016 & -0.069 & -0.068 & 18.068 & 18.137 & 18.205 & 17.750 & 17.774 & 17.992 &  37.32  \\
1992bl & -0.032 & -0.009 & -0.192 & 17.411 & 17.420 & 17.613 & 17.119 & 17.143 & 17.399 &  36.69  \\
1992bo & -0.007 &  0.000 & -0.136 & 15.774 & 15.774 & 15.911 & 15.396 & 15.420 & 15.632 &  34.97  \\
1992bp & -0.023 & -0.046 & -0.210 & 18.330 & 18.377 & 18.587 & 18.089 & 18.113 & 18.421 &  37.66  \\
1992br &  0.010 &  0.034 &\nodatr & 19.196 & 19.162 &\nodatr & 18.795 & 18.819 &\nodatr &  38.37  \\
1992bs &  0.014 & -0.025 &\nodatr & 18.234 & 18.259 &\nodatr & 18.204 & 18.228 &\nodatr &  37.78  \\
1993B  &  0.054 & -0.073 & -0.234 & 18.188 & 18.261 & 18.494 & 18.099 & 18.123 & 18.447 &  37.67  \\
1993H  &  0.122 &  0.088 & -0.119 & 16.299 & 16.211 & 16.327 & 15.860 & 15.884 & 15.976 &  35.43  \\
1993L  &  0.193 & -0.027 &\nodatr & 12.640 & 12.666 &\nodatr & 12.415 & 12.439 &\nodatr &  31.99  \\
1993O  & -0.031 & -0.092 & -0.158 & 17.685 & 17.777 & 17.936 & 17.635 & 17.659 & 17.922 &  37.21  \\
1993ac &  0.040 & -0.033 & -0.135 & 17.606 & 17.639 & 17.773 & 17.520 & 17.544 & 17.714 &  37.09  \\
1993ae & -0.057 & -0.081 & -0.105 & 15.471 & 15.552 & 15.658 & 15.284 & 15.308 & 15.543 &  34.86  \\
1993ag &  0.044 &  0.024 & -0.303 & 17.682 & 17.657 & 17.960 & 17.526 & 17.550 & 17.832 &  37.10  \\
1993ah & -0.023 & -0.037 & -0.286 & 16.322 & 16.359 & 16.645 & 16.117 & 16.141 & 16.502 &  35.69  \\
1994D  & -0.040 & -0.062 & -0.146 & 11.916 & 11.978 & 12.124 & 11.814 & 11.838 & 12.064 &  31.39  \\
1994M  &  0.056 & -0.029 & -0.204 & 16.050 & 16.079 & 16.282 & 15.839 & 15.863 & 16.133 &  35.41  \\
1994Q  &  0.035 &  0.028 & -0.338 & 16.243 & 16.215 & 16.552 & 16.329 & 16.353 & 16.597 &  35.90  \\
1994S  & -0.042 & -0.019 & -0.281 & 14.858 & 14.877 & 15.159 & 14.904 & 14.928 & 15.190 &  34.48  \\
1994T  &  0.047 &  0.044 & -0.250 & 17.058 & 17.014 & 17.264 & 16.797 & 16.821 & 17.054 &  36.37  \\
1994ae &  0.033 & -0.013 & -0.382 & 12.907 & 12.920 & 13.302 & 13.040 & 13.064 & 13.393 &  32.61  \\
1995D  & -0.009 & -0.009 & -0.364 & 13.234 & 13.243 & 13.607 & 13.267 & 13.291 & 13.626 &  32.84  \\
1995E  &  0.647 &  0.046 &\nodatr & 14.347 & 14.301 &\nodatr & 14.244 & 14.268 &\nodatr &  33.82  \\
1995ak &  0.055 & -0.103 & -0.232 & 15.734 & 15.837 & 16.068 & 15.574 & 15.598 & 15.979 &  35.15  \\
1995al &  0.071 &  0.025 & -0.362 & 13.043 & 13.018 & 13.379 & 13.138 & 13.162 & 13.431 &  32.71  \\
1996C  &  0.026 & -0.009 & -0.262 & 16.440 & 16.450 & 16.711 & 16.528 & 16.552 & 16.769 &  36.10  \\
1996X  & -0.015 & -0.004 & -0.250 & 13.032 & 13.036 & 13.286 & 12.883 & 12.907 & 13.173 &  32.46  \\
1996Z  &  0.332 & -0.036 &\nodatr & 13.135 & 13.172 &\nodatr & 13.070 & 13.094 &\nodatr &  32.64  \\
1996ab &  0.101 & -0.013 &\nodatr & 19.040 & 19.053 &\nodatr & 19.173 & 19.197 &\nodatr &  38.75  \\
1996ai &  1.611 &  0.075 &\nodatr & 11.023 & 10.948 &\nodatr & 11.107 & 11.131 &\nodatr & (30.68) \\
1996bk &  0.308 &  0.104 &\nodatr & 13.641 & 13.538 &\nodatr & 13.192 & 13.216 &\nodatr &  32.77  \\
1996bl &  0.042 & -0.034 & -0.225 & 16.550 & 16.584 & 16.808 & 16.551 & 16.575 & 16.812 &  36.13  \\
1996bo &  0.257 &  0.036 &\nodatr & 14.898 & 14.862 &\nodatr & 14.734 & 14.758 &\nodatr &  34.31  \\
1996bv &  0.117 &  0.008 & -0.311 & 14.913 & 14.905 & 15.213 & 15.050 & 15.074 & 15.301 &  34.62  \\
1997E  &  0.024 & -0.008 & -0.202 & 14.996 & 15.003 & 15.205 & 14.807 & 14.831 & 15.064 &  34.38  \\
1997Y  &  0.035 & -0.072 & -0.128 & 15.084 & 15.156 & 15.283 & 15.025 & 15.049 & 15.256 &  34.60  \\
1997bp &  0.169 & -0.013 & -0.451 & 13.322 & 13.335 & 13.782 & 13.394 & 13.418 & 13.831 &  32.97  \\
1997bq &  0.159 &  0.117 & -0.341 & 13.890 & 13.774 & 14.112 & 13.848 & 13.872 & 14.035 &  33.42  \\
1997cw &  0.334 &  0.073 & -0.346 & 14.481 & 14.408 & 14.748 & 14.463 & 14.487 & 14.703 & (34.04) \\
1997dg &  0.057 & -0.045 & -0.207 & 16.670 & 16.716 & 16.922 & 16.667 & 16.691 & 16.927 &  36.24  \\
1997do &  0.055 & -0.018 & -0.284 & 14.103 & 14.120 & 14.403 & 14.166 & 14.190 & 14.446 &  33.74  \\
1997dt &  0.496 &  0.007 & -0.189 & 13.595 & 13.588 & 13.768 & 13.611 & 13.635 & 13.768 &  33.19  \\
1998V  &  0.016 & -0.042 & -0.186 & 15.019 & 15.060 & 15.246 & 15.055 & 15.079 & 15.278 &  34.63  \\
1998ab &  0.079 & -0.086 & -0.177 & 15.581 & 15.667 & 15.842 & 15.759 & 15.783 & 15.990 & (35.33) \\
1998aq & -0.052 & -0.022 &\nodatr & 12.353 & 12.375 &\nodatr & 12.339 & 12.363 &\nodatr &  31.91  \\
1998bu &  0.279 &  0.056 & -0.191 & 11.098 & 11.043 & 11.228 & 11.013 & 11.037 & 11.140 &  30.59  \\
1998dh &  0.120 &  0.072 & -0.307 & 13.522 & 13.451 & 13.755 & 13.377 & 13.401 & 13.619 &  32.95  \\
1998dk &  0.130 &  0.016 & -0.299 & 14.275 & 14.259 & 14.555 & 14.278 & 14.302 & 14.544 &  33.85  \\
1998dm &  0.239 & -0.063 & -0.173 & 13.646 & 13.710 & 13.878 & 13.692 & 13.716 & 13.923 &  33.27  \\
1998dx & -0.026 & -0.045 & -0.139 & 17.638 & 17.683 & 17.822 & 17.377 & 17.401 & 17.641 &  36.95  \\
1998ec &  0.131 &  0.014 & -0.293 & 15.614 & 15.600 & 15.890 & 15.600 & 15.624 & 15.868 &  35.17  \\
1998ef &  0.021 & -0.064 & -0.223 & 14.833 & 14.897 & 15.120 & 14.941 & 14.965 & 15.210 &  34.52  \\
1998eg &  0.043 & -0.046 & -0.206 & 15.959 & 16.005 & 16.211 & 15.944 & 15.968 & 16.207 &  35.52  \\
1998es &  0.055 &  0.033 & -0.354 & 13.660 & 13.626 & 13.979 & 13.761 & 13.785 & 14.032 & (33.34) \\
1999X  &  0.139 & -0.011 & -0.245 & 15.812 & 15.823 & 16.065 & 15.797 & 15.821 & 16.050 &  35.37  \\
1999aa & -0.022 &  0.022 & -0.382 & 14.838 & 14.815 & 15.198 & 14.959 & 14.983 & 15.270 & (34.53) \\
1999ac &  0.077 & -0.063 & -0.202 & 13.869 & 13.932 & 14.133 & 13.957 & 13.981 & 14.209 & (33.53) \\
1999aw & -0.045 &  0.033 & -0.382 & 16.893 & 16.860 & 17.243 & 17.031 & 17.055 & 17.323 &  36.61  \\
1999cc & -0.002 &  0.009 & -0.237 & 16.763 & 16.754 & 16.991 & 16.520 & 16.544 & 16.806 &  36.09  \\
1999cl &  0.918 &  0.264 & -0.451 & 11.605 & 11.341 & 11.773 & 11.351 & 11.375 & 11.495 & (30.93) \\
1999dk &  0.024 &  0.012 & -0.372 & 14.732 & 14.720 & 15.091 & 14.597 & 14.621 & 14.982 &  34.17  \\
1999dq &  0.096 & -0.006 & -0.350 & 14.078 & 14.084 & 14.432 & 14.200 & 14.224 & 14.514 & (33.77) \\
1999ee &  0.234 &  0.066 & -0.385 & 13.992 & 13.927 & 14.307 & 14.040 & 14.064 & 14.312 &  33.61  \\
1999ef & -0.010 &  0.053 & -0.399 & 17.201 & 17.148 & 17.547 & 17.172 & 17.196 & 17.501 &  36.75  \\
1999ej &  0.015 & -0.036 & -0.172 & 15.304 & 15.340 & 15.512 & 15.123 & 15.147 & 15.386 & (34.70) \\
1999ek &  0.177 &  0.042 & -0.313 & 15.023 & 14.981 & 15.291 & 14.959 & 14.983 & 15.223 &  34.53  \\
1999gd &  0.366 &  0.003 & -0.197 & 15.515 & 15.513 & 15.702 & 15.472 & 15.496 & 15.662 &  35.05  \\
1999gh &  0.084 &  0.048 & -0.166 & 13.917 & 13.869 & 14.033 & 13.506 & 13.530 & 13.714 &  33.08  \\
1999gp &  0.027 &  0.067 & -0.438 & 15.922 & 15.855 & 16.293 & 15.957 & 15.981 & 16.288 & (35.53) \\
2000B  &  0.036 &  0.066 & -0.306 & 15.530 & 15.464 & 15.769 & 15.248 & 15.272 & 15.537 &  34.82  \\
2000E  &  0.136 & -0.022 & -0.315 & 12.313 & 12.335 & 12.648 & 12.410 & 12.434 & 12.716 &  31.98  \\
2000bk &  0.099 & -0.094 & -0.023 & 16.518 & 16.611 & 16.632 & 16.205 & 16.229 & 16.430 &  35.78  \\
2000ce &  0.490 &  0.103 & -0.281 & 15.287 & 15.184 & 15.456 & 15.261 & 15.285 & 15.395 &  34.84  \\
2000cf &  0.013 &  0.005 & -0.269 & 16.971 & 16.966 & 17.235 & 16.847 & 16.871 & 17.136 &  36.42  \\
2000cn &  0.033 &  0.100 & -0.197 & 16.467 & 16.366 & 16.563 & 16.087 & 16.111 & 16.250 &  35.66  \\
2000dk & -0.036 &  0.036 & -0.224 & 15.473 & 15.438 & 15.662 & 15.144 & 15.168 & 15.406 &  34.72  \\
2000fa &  0.018 & -0.047 & -0.284 & 15.640 & 15.687 & 15.971 & 15.717 & 15.741 & 16.034 &  35.29  \\
2001V  &  0.032 & -0.012 & -0.308 & 14.443 & 14.454 & 14.761 & 14.502 & 14.526 & 14.799 &  34.08  \\
2001el &  0.128 & -0.062 & -0.277 & 12.285 & 12.347 & 12.622 & 12.293 & 12.317 & 12.640 &  31.87  \\
2002bo &  0.316 &  0.119 & -0.390 & 12.783 & 12.665 & 13.048 & 12.666 & 12.690 & 12.917 &  32.24  \\
2002er &  0.117 &  0.026 & -0.259 & 13.819 & 13.793 & 14.049 & 13.644 & 13.668 & 13.907 &  33.22  \\
\tableline
\noalign{\smallskip}
\multicolumn{11}{c}{\small (b) SNe\,Ia-T}\\
\noalign{\smallskip}
\tableline
1991T  &  0.143 &  0.025 & -0.338 & 11.077 & 11.052 & 11.387 & 11.141 & 11.165 & 11.417 & \\
1995ac &  0.020 & -0.062 & -0.231 & 16.965 & 17.027 & 17.257 & 17.083 & 17.107 & 17.355 & \\
1995bd &  0.305 & -0.033 &\nodatr & 14.113 & 14.147 &\nodatr & 14.255 & 14.279 &\nodatr & \\
1997br &  0.252 & -0.005 & -0.240 & 12.656 & 12.661 & 12.896 & 12.686 & 12.710 & 12.911 & \\
\noalign{\smallskip}
2000cx &  0.007 &  \multicolumn{9}{l}{\hspace*{0.8cm}see Table~6}  \\
\tableline
\noalign{\smallskip}
\multicolumn{11}{c}{\small (c) SNe\,Ia-bg}\\
\noalign{\smallskip}
\tableline
1991bg & 0.00\tablenotemark{1)} &  0.749 &  0.397 & 14.582 & 13.833 & 13.436 & & & & \\
1992K  & 0.00\tablenotemark{1)} &  0.789 &  0.129 & 15.896 & 15.107 & 14.978 & & & & \\
1997cn & 0.00\tablenotemark{1)} &  0.513 &  0.215 & 16.819 & 16.306 & 16.091 & & & & \\
1998bp & 0.00\tablenotemark{1)} &  0.404 &  0.141 & 15.418 & 15.014 & 14.873 & & & & \\
1998de & 0.00\tablenotemark{1)} &  0.702 &  0.245 & 17.322 & 16.620 & 16.376 & & & & \\
1999by & 0.00\tablenotemark{1)} &  0.494 &  0.219 & 13.594 & 13.100 & 12.881 & & & & \\
1999da & 0.00\tablenotemark{1)} &  0.682 &  0.235 & 16.662 & 15.980 & 15.746 & & & & \\
\noalign{\smallskip}                                                                
1986G  & \multicolumn{9}{l}{\hspace*{0.8cm}see Table~7}   \\
\enddata
\tablenotetext{1)}{$E(B\!-\!V)_{\rm host}$ assumed to be 0.00 (see
  \S~\ref{sec:SNpec:bg}).} 
\tablecomments{Negative extinction and absorption corrections in the
  host galaxies have been subtracted algebraicly in columns 3-9 in
  order not to skew the distribution function and not to shift the
  {\em mean\/} value. For the most probable values of {\em
  individual\/} SNe\,Ia negative extinctions $E(B\!-\!V)_{\rm host}$
  and corresponding absorption corrections should be set to zero.}
\end{deluxetable}

\end{document}